\documentclass[aps,showpacs,prl,twocolumn,superscriptaddress,floatfix]{revtex4-2}

\usepackage[T1]{fontenc}
\usepackage{textcomp}

\usepackage{graphicx, bm,amssymb,amsmath,dcolumn}
\usepackage{multirow}
\usepackage{enumerate}
\usepackage{enumitem}
\usepackage{color}
\usepackage{subfigure}  
\usepackage{epstopdf}
\usepackage{bbm}
\usepackage{threeparttable}
\usepackage{amsthm}
\usepackage{mathtools}
\usepackage{tikz}
\usepackage{float}
\usepackage{empheq}
\usepackage{algorithm}
\usepackage{algpseudocode}
\usepackage{array}
\usepackage{tabularx}
\usepackage{pifont}
\usepackage{hyperref}

\begin{document}
\title{Finite-Temperature Phase Transition and Logarithmic Quasi-Long-Range Order \\ 
in the Two-dimensional Long-range Heisenberg Model}

\author{Tianning Xiao}
\thanks{These two authors contributed equally to this work}
\affiliation{Hefei National Research Center for Physical Sciences at the Microscale and School of Physical Sciences, University of Science and Technology of China, Hefei 230026, China}

\author{Dingyun Yao}
\thanks{These two authors contributed equally to this work}
\affiliation{Hefei National Research Center for Physical Sciences at the Microscale and School of Physical Sciences, University of Science and Technology of China, Hefei 230026, China}

\author{Lode Pollet}
\affiliation{Department of Physics and Arnold Sommerfeld Center for Theoretical Physics (ASC), Ludwig-Maximilians-Universit{\"a}t M{\"u}nchen, Theresienstrasse 37, M{\"u}nchen D-80333, Germany}
\affiliation{Munich Center for Quantum Science and Technology (MCQST), Schellingstrasse 4, D-80799 M{\"u}nchen, Germany}

\author{Zhijie Fan}
\email{zfanac@ustc.edu.cn}
\affiliation{Hefei National Research Center for Physical Sciences at the Microscale and School of Physical Sciences, University of Science and Technology of China, Hefei 230026, China}
\affiliation{Hefei National Laboratory, University of Science and Technology of China, Hefei 230088, China}
\affiliation{Shanghai Research Center for Quantum Science and CAS Center for Excellence in Quantum Information and Quantum Physics, University of Science and Technology of China, Shanghai 201315, China}

\author{Youjin Deng}
\email{yjdeng@ustc.edu.cn}
\affiliation{Hefei National Research Center for Physical Sciences at the Microscale and School of Physical Sciences, University of Science and Technology of China, Hefei 230026, China}
\affiliation{Hefei National Laboratory, University of Science and Technology of China, Hefei 230088, China}
\affiliation{Shanghai Research Center for Quantum Science and CAS Center for Excellence in Quantum Information and Quantum Physics, University of Science and Technology of China, Shanghai 201315, China}

\begin{abstract}
We investigate the two-dimensional classical Heisenberg ($N=3$) model with long-range (LR)
interactions decaying as \(1/r^{2+\sigma}\), focusing on the marginal case
\(\sigma=2\). Large-scale Monte Carlo simulations up to linear size \(L=8192\)
reveal a finite-temperature phase transition, in striking contrast to the
short-range case where the system is asymptotically free. An adiabatic
renormalization-group analysis shows that the LR interaction provides a
positive contribution to the stiffness flow and gives rise to a weak continuous transition
through non-topological spin-wave fluctuations. We show that the 
low-temperature phase exhibits a novel logarithmic quasi-long-range order
(log-QLRO), with correlations decaying as \(C(r)\sim(\ln r)^{-\eta_\ell}\) and
vanishing magnetization in the strict thermodynamic limit. High-precision data
confirm this picture, including the predicted identity
\(2\pi\kappa_R\eta_\ell=N-1\), with \(\kappa_R\) the renormalized spin stiffness.
The decay, with \(\eta_\ell\propto T\), is slower than any power law, making the phase
practically indistinguishable from true long-range order in realistic systems.
Our results establish a novel log-QLRO phase and a continuous transition beyond
the Ginzburg-Landau-Wilson paradigm in the two-dimensional Heisenberg
model at \(\sigma=2\).
\end{abstract}
\maketitle

Low-dimensional systems have played a central role in condensed-matter and statistical physics, with examples ranging from the quantum Hall effect~\cite{RevModPhys.58.519, PhysRevLett.48.1559, doi:10.1126/science.1234414} to topological insulators~\cite{PhysRevLett.95.226801, doi:10.1126/science.1133734, doi:10.1126/science.1148047}. 
For systems with short-range (SR) interactions and continuous symmetries in dimensions $d \leq 2$, 
the Mermin--Wagner theorem excludes spontaneous symmetry breaking at finite temperature (finite-\(T\))~\cite{PhysRevLett.17.1133,merminAbsenceOrderingCertain1967,Pfister1981Symmetry}.
Nevertheless, a finite-\(T\) transition without long-range order (LRO) can still occur~\cite{Shlosman1980PhaseTransitions,vanEnterShlosman2005}, as in the Berezinskii--Kosterlitz--Thouless (BKT) transition of two-dimensional (2D) XY models~\cite{kosterlitz2017}.

Introducing long-range (LR) interactions like $1/r^{d+\sigma}$ can qualitatively change this picture by stabilizing LRO. 
Early renormalization-group (RG) analyses by Fisher \emph{et al.} identified a crossover threshold $\sigma_* = 2$ for LR $O(N)$ models~\cite{Fisher1972}, while Sak later proposed $\sigma_* = 2 - \eta_{\mathrm{SR}}$, with $\eta_{\mathrm{SR}}$ the anomalous dimension of the SR fixed point~\cite{Sak1973}. This widely accepted criterion has recently been challenged by numerical and theoretical studies, leaving the LR--SR boundary under debate~\cite{Picco2012, Blanchard2013, Grassberger2013, Xiao2025LRXY, Yao2025LRXY, Liu2025Percolation1, Xiao2025Ising, Li2026Ising,Li2026Quantum,Li2026Percolation}.

In 2D LR systems with continuous symmetry, the extended Mermin-Wagner theorem~\cite{Joyce1969, Pfister1981Symmetry, Bruno2001} shows that finite-\(T\) LRO is absent for $\sigma \geq 2$. The XY model ($N=2$) phase diagram is complicated by BKT physics: some propose an intermediate BKT phase for $7/4 \leq \sigma < 2$~\cite{Giachetti2021}, while others argue that the BKT transition persists for all $\sigma>0$~\cite{Walther2026}. 
High-precision simulations instead found a single continuous transition into an LRO phase 
for $\sigma < 2$~\cite{Xiao2025LRXY, Yao2025LRXY}.
Field-theoretic treatments also indicate that the SR Wilson-Fisher fixed point becomes unstable for $\sigma < 2$~\cite{Li2026Ising,Li2026Quantum,Li2026Percolation}.
These controversies, together with experimental realizations of LR interactions in quantum simulators~\cite{Monroe2013, Browaeys2023}, motivate a cleaner test.

The natural test is its \(O(3)\) counterpart.
The 2D SR Heisenberg ($N=3$) model is asymptotically free: spin-wave interactions soften the spin stiffness, and no finite-\(T\) transition exists~\cite{NLSM_O3_2,NLSM_O3_1,PhysRevB.111.214403}.
Unlike the XY model, it does not support topologically stable vortices, since its order-parameter space $S^2$ is simply connected~\cite{Mermin1979Defects}.
Therefore, if a finite-\(T\) transition appears at \(\sigma=2\), it must arise from a purely non-topological mechanism.

\begin{figure}[t]
    \centering
    \includegraphics[width=0.9\linewidth]{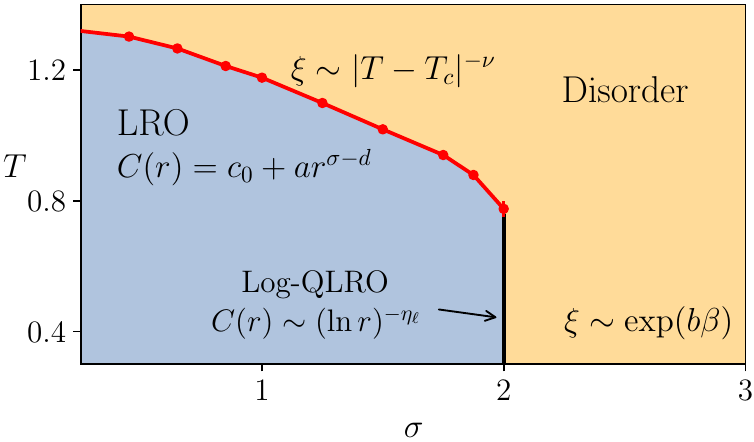}
    \vspace{-10pt}
    \caption{Phase diagram of the 2D LR-Heisenberg model.
    The solid red line is the finite-\(T\) transition $T_c(\sigma)$. The low-T phase is LRO for $\sigma<2$ and log-QLRO at $\sigma=2$;  no finite-\(T\) transition exists for $\sigma>2$. Labels summarize the asymptotic behavior of the correlation function \(C(r)\) and the correlation length \(\xi\).
    }
    \label{PD}
    \vspace{-15pt}
\end{figure}

We perform large-scale Monte Carlo simulations of the 2D LR-Heisenberg model and obtain the phase diagram in Fig.~\ref{PD}. The model undergoes a single continuous transition for all \(\sigma \leq 2\).
For \(\sigma>2\), it recovers the SR behavior, and no finite-\(T\) transition exists. 
For \(\sigma<2\), \(T_c\) increases monotonically as \(\sigma\) decreases;
the low-T phase is genuinely LRO, and the correlation function saturates to a constant  as \(C(r) \simeq c_0 + a r^{\sigma-d}\), consistent with a Gaussian long-wavelength description of the transverse (Goldstone) fluctuations.
At \(\sigma=2\), the transition occurs at a finite temperature, with \(\beta_c = 1.29(2)\) (\(T_c \approx 0.78\)). 

To understand this marginal behavior at \(\sigma=2\), we outline an RG analysis within the adiabatic approximation. At the one-loop level in the nonlinear sigma model (NLSM) 
for the low-energy spin waves, we define an effective spin stiffness
\(K(\ell)=K_0+(\kappa-\kappa_c)\ell\), where \(K_0\) is the bare SR
stiffness, \( \kappa\) is the dimensionless LR coupling strength,
\(\kappa_c=(N-2)/(2\pi)\), and \(\ell=\ln(\Lambda/\mu)\) is the logarithmic
RG scale. In the SR limit \(\kappa=0\),
\(K(\ell)=K_0-\kappa_c\ell\) decreases linearly with \(\ell\), reflecting
the asymptotic freedom. The LR interaction, however, contributes
a positive constant \(\kappa \propto \beta \) to the stiffness flow \(dK/d\ell\). 
Setting $\kappa=\kappa_c$ predicts a critical inverse temperature $\beta_c$, 
and further, $\beta_c \propto N$ for large $N$.
Higher-loop corrections shift the actual transition to \(\kappa_c^*>\kappa_c\),
with the inverse correlation-length exponent 
\(1/\nu \propto (\kappa_c^*-\kappa_c)^{2}\). This suggests a weak
transition with a large \(\nu\), consistent with our numerical estimate
\(1/\nu=0.13(2)\). 
For \(N=1,2,3\), \(\beta_c=0.344439(2)\)~\cite{Xiao2025Ising},  
\(0.7315(2)\)~\cite{Xiao2025LRXY,Yao2025LRXY}, and \(1.29(2)\), respectively, 
which satisfy an approximately linear fit \(\beta_c\approx-0.157+0.473N\), 
although it is far from the asymptotic limit.

At \(\sigma=2\), the effective stiffness in the low-T phase grows
logarithmically in the infrared, with the leading kernel
\(\gamma(k) \propto k^2 \ln(1/k)\). The spin-wave theory
predicts a novel logarithmic quasi-long-range order (log-QLRO) phase with unusual scaling behaviors. 
The correlation function decays as \(C(r) \sim (\ln r)^{-\eta_\ell}\), slower than any power law; 
the  squared magnetization vanishes as \(\langle M^2 \rangle \sim (\ln L)^{-\eta_\ell}\); 
the correlation-length ratio diverges as \((\xi/L)^2 \simeq s \ln L\).
Here,  $\eta_\ell = (N-1)/(2\pi \kappa_R)$ and $s=\kappa_R/(N-1)$,
with \(\kappa_R\) the renormalized spin stiffness 
and \( (N-1)\) the number of transverse modes.
High-precision data confirm these characteristic features, 
including the identity \(2\pi\kappa_R\eta_\ell=N-1\).
Despite the absence of a true LRO in the strict thermodynamic limit, the extremely slow decay makes the
log-QLRO phase practically indistinguishable from true LRO in realistic
finite systems, including materials with Avogadro-scale atom numbers.
Notably, this log-QLRO phase and the associated continuous transition
transcend the Ginzburg-Landau-Wilson (GLW) paradigm: the low-T phase
is not described by a local order parameter, and the 
exponent \(\nu\) is non-universal, depending explicitly on \(\kappa\).

\begin{table}[t]
    \centering
    \caption{
    Critical points $\beta_c$ and critical exponents for representative values of $\sigma$ in the 2D LR-Heisenberg model, where $\nu$ is the correlation-length exponent, and $\eta$ is the anomalous dimension.
    }
    \begin{tabular}{D{.}{.}{-1}|D{.}{.}{-1}D{.}{.}{-1}D{.}{.}{-1}}
    \hline\hline
    \multicolumn{1}{c|}{$\sigma$} &
    \multicolumn{1}{c}{$\beta_c$} &
    \multicolumn{1}{c}{$1/\nu$} &
    \multicolumn{1}{c}{$\eta$} \\
    \hline
    1.75  & 1.0641(4) & 0.34(3) & 0.295(5) \\
    1.875 & 1.138(4)  & 0.21(2) & 0.192(6) \\
    2     & 1.29(2)   & 0.13(2) & 0.132(4) \\
    \hline\hline
    \end{tabular}
    \label{tab:critical_parameters}
    \vspace{-10pt}
\end{table}

{\it Model and observables} --- 
We simulate the LR Heisenberg model, ${\mathcal H}=-\sum_{i<j} f(r_{ij}) \bm S_i\cdot\bm S_j$, on a periodic square lattice of linear size $L$, with $\bm S_i$ a three-component unit vector.
The constant $b$, in $f(r) = b/r^{d+\sigma}$, is fixed by $4 = \sum_{j \neq 0} f(r_{0j})$~\cite{Xiao2025LRXY,Yao2025LRXY}, so the energy per site of a fully polarized configuration is \(-2\), independent of \(L\). This normalization reduces finite-size corrections by suppressing the \(L\)-dependent drift of the background energy density.
We employ an enhanced Luijten-Bl\"ote (LB) cluster algorithm~\cite{luijten1995,luijten1997}, which significantly accelerates cluster construction for LR interactions and enables simulations up to $L=8192$ (see details in Ref.~\cite{Yao2025LRXY}). 
We measure the squared magnetization $M^2=\left|L^{-d}\sum_i\bm S_i\right|^2$ and its Fourier mode $M_k^2=\left|L^{-d}\sum_i\bm S_i e^{i\bm k\cdot\bm r_i}\right|^2$, with $\bm k=(2\pi/L,0)$. 
We define the susceptibility $\chi=L^d\langle M^2\rangle$, the Fourier susceptibility $\chi_k=L^d\langle M_k^2\rangle$, and the second-moment correlation length $\xi=[2\sin(|\bm k|/2)]^{-1}\sqrt{\chi/\chi_k-1}$. 

\begin{figure}[t]
    \centering
    \includegraphics[width=\linewidth]{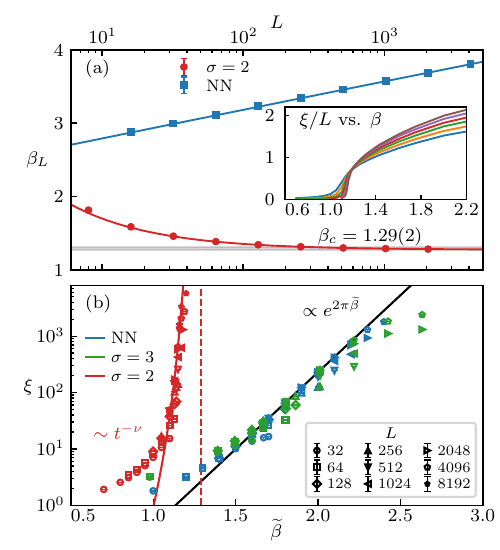}
    \vspace{-15pt}
    \caption{Finite-temperature phase transition of the 2D LR-Heisenberg model at
$\sigma=2$. (a) Pseudocritical point $\beta_L$, defined by $\xi(\beta_L)/L=1$,
decreases monotonically with increasing $L$ and converges to a finite
$\beta_c=1.29(2)$ (gray band), in contrast to the nearest-neighbor (NN) limit
$\sigma\to\infty$, where $\beta_L\simeq(1/2\pi)\ln L$ diverges logarithmically.
Inset: $\xi/L$ versus $\beta$ for $L\in[64,2048]$, with the crossing near
$\beta_c\approx1.3$. (b) Correlation length $\xi$ versus
$\tilde{\beta}$, where $\tilde{\beta}=2.07\beta-1.09$ for $\sigma=3$ and
$\tilde{\beta}=\beta$ otherwise. Data for $\sigma>2$ collapse onto the
exponential growth $\xi\propto e^{2\pi\tilde{\beta}}$ (black line), whereas for
$\sigma=2$, $\xi$ diverges at $\beta_c$ (dashed line) as $\xi\sim t^{-\nu}$
with $t\equiv\beta_c-\beta$ and $\nu\approx8$ (red curve).}
    \label{fig:sigma2}
    \vspace{-15pt}
\end{figure}

{\it Finite-temperature transition at $\sigma=2$} ---
We first examine $\sigma=2$, where the leading kernel is $\gamma(k) \sim k^2 \ln (1/k)$. 
Figure~\ref{fig:sigma2}(a) shows a finite-\(T\) transition 
from the pseudocritical point $\beta_L$, defined by $\xi(\beta_L)/L=1$,
which decreases with increasing $L$ and converges to $\beta_c=1.29(2)$ as $\beta_L=\beta_c+bL^{-0.73}$. 
The approximate value of $\beta_c$ is readily identified by the crossing of 
the curves of $\xi/L$ (inset of Fig.~\ref{fig:sigma2}(a)).
Fitting details are in the Supplemental Material (SM)~\cite{SM}.
In contrast, in the nearest-neighbor (NN) limit, $\beta_L$ diverges as $\beta_L\simeq (1/2\pi)\ln L$, illustrating asymptotic freedom~\cite{PhysRevB.111.214403}.

Figure~\ref{fig:sigma2}(b) provides further evidence for a second-order transition from the growth of $\xi$ on the high-T side. 
Because finite $L$ cuts off $\xi$ at $\mathcal O(L)$, we progressively increase $L$ with $\beta$ to track the growth of $\xi$~\cite{PhysRevB.111.214403}. 
For the NN case, $\xi$ is finite at any finite $T$ and grows as $\xi\sim e^{2\pi\beta}$ as $ \beta \to \infty$~\cite{CARACCIOLO1994141, FALCIONI1986671, PhysRevB.54.990, PhysRevB.111.214403}; for $\sigma=3$, the rescaling $\beta\rightarrow a\beta+b$ collapses the data onto the NN curve.
For \(\sigma=2\), however, the data bend upward in the semi-log plot and depart from the SR form,
consistent with the power-law divergence, $\xi\sim t^{-\nu}$, with $t\equiv\beta_c-\beta$ and $\nu\approx8$.

At \(\sigma=2\), \(\beta_L\) defined by \(\xi(\beta_L)/L=1\) approaches \(\beta_c\) from the log-QLRO side, 
and the drift is described by \(\beta_L=\beta_c+bL^{-0.73}\);
if there were systematic bias, a true \(\beta_c\) would be slightly below \(1.29(2)\).
We further perform standard FSS of the \(\xi/L\) data near criticality 
and obtain only an effective \(y_t\equiv1/\nu\approx0.21\)--\(0.29\) (see SM), 
due to strong finite-size corrections and the nearby log-QLRO phase.
Both the drift exponent \(0.73\) and the fit value \(0.21\)--\(0.29\)
are therefore effective, not asymptotic estimates of \(1/\nu\).
A reliable determination proceeds in two steps: first locate \(\beta_c\)
from the crossing of \(\xi/L\), then perform FSS at \(\beta_c\) using the
energy-magnetization covariance \(C_{em}\), whose singular part scales as
\(L^{y_t}\), as in Refs.~\cite{Xiao2025LRXY,Yao2025LRXY}. This gives
\(1/\nu=0.13(2)\) in Table~\ref{tab:critical_parameters}, consistent
with \(\nu\approx8\) in Fig.~\ref{fig:sigma2}(b). The anomalous
dimension \(\eta\) is obtained from the FSS of \(\chi\) at \(\beta_c\).

\begin{figure}[t]
\centering
\includegraphics[width=\linewidth]{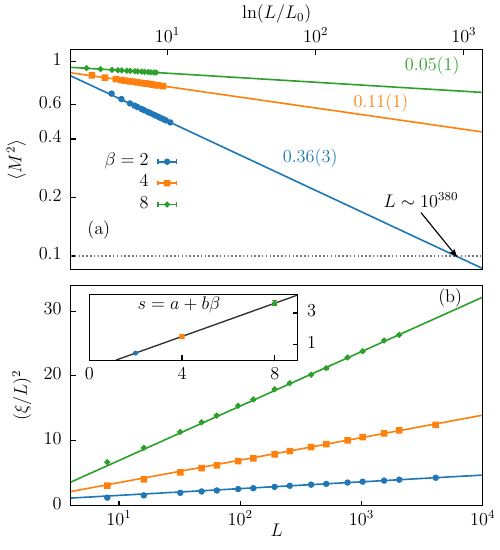}
\vspace{-20pt}
\caption{(a) Log-log plot of \(\langle M^2\rangle\) versus \(\ln(L/L_0)\). The linear behavior yields \(\langle M^2\rangle \sim (\ln L)^{-\eta_\ell}\), which decays slower than any power law but still vanishes in the strict thermodynamic limit.
For \(\beta=2\), reaching \(\langle M^2\rangle\approx0.1\) would require \(L\approx10^{380}\), illustrating the practical consequence of the exceptionally slow decay.
(b) Semi-log plot of \( (\xi/L)^2\) versus \(L\). The linear behavior confirms \((\xi/L)^2 \sim s \ln L\), and the slope follows \(s =a+b\beta\) (inset), directly demonstrating the logarithmic stiffness kernel. 
}
\label{fig:logqlro}
\vspace{-15pt}
\end{figure}

{\it Renormalization-group and spin-wave analyses} --- 
The low-energy physics of the 2D \(O(N)\) spin model for $N>2$ is described by the nonlinear sigma model (NLSM). For \(\sigma=2\), the action in momentum space reads
\begin{equation}
S=\frac{1}{2}\int\frac{d^2k}{(2\pi)^2}
\left[K_0 k^2+\kappa k^2\ln\frac{1}{k}\right]
|\mathbf{n}_k|^2,
\label{eq:action}
\end{equation}
where \(\mathbf{n}_k\) is the Fourier transform of the unit spin field, and \(K_0\) and \(\kappa\) are respectively the dimensionless SR and LR coupling strengths. Within the adiabatic approximation, we absorb the LR contribution into a running effective spin stiffness,
which at the one-loop level reads
\begin{equation}
K(\ell)=K_0+\kappa\ell-\frac{N-2}{2\pi}\ell
=K_0+\left(\kappa-\kappa_c\right)\ell.
\label{eq:Krunning}
\end{equation}
Here, \( \kappa_c=(N-2)/2\pi\) and \(\ell=\ln(\Lambda/\mu)\) is the logarithmic RG scale, with \(\Lambda\) the ultraviolet cutoff and \(\mu\) the renormalization scale. 
The adiabatic approximation is exact at the Gaussian level,
and the approximation enters in assuming that \(\kappa\) itself does not flow significantly and is thus fixed as a constant. Equivalently, the RG flow of the stiffness is
\begin{equation}
\frac{dK}{d\ell}=\kappa-\kappa_c,
\label{eq:Kflow}
\end{equation}
where the first term is the LR enhancement and the second term is the SR softening. In the SR limit \(\kappa=0\), the asymptotic freedom is recovered. 
Note that, as temperature decreases, $\kappa \propto \beta$ increases while $\kappa_c$ remains constant.
When \(\kappa<\kappa_c\), the stiffness is driven to zero in the infrared, leading to a disordered state. When \(\kappa>\kappa_c\), the LR enhancement wins, and $K$ grows linearly in the logarithmic RG scale \(\ell\).

\begin{table}[t]
\centering
\caption{Log-QLRO parameters at \(\sigma=2\), with \(\eta_{\ell}\) and
\(\kappa_{\mathrm{R}}\) extracted respectively from \(\langle M^{2}\rangle\)
and \((\xi/L)^{2}\). The column \(\kappa_{\rm 1L}\) is the one-loop reference.
The last column tests the spin-wave relation
\(\mathcal{P}=2\pi\kappa_{\mathrm{R}}\eta_{\ell}=N-1\).}
\label{tab:logqlro}
\begin{tabular}{lllll}
\hline \hline 
\(\beta\) & \(\eta_{\ell}\) & \(\kappa_R\) &
\(\kappa_{\rm 1L} \) & \(\mathcal{P}\) \\
\hline 
2 & 0.36(3)  & 0.90(2)      & 1.93    & 2.03(20)  \\
4 & 0.11(1)  & 3.00(9)\;    & 4.01    & 2.07(20) \\
8\;\; & 0.05(1)\;  & 7.3(3) & 8.18 \; & 2.30(50) \\
\hline \hline 
\end{tabular}
\end{table}

Higher-loop corrections shift the actual transition to a larger value $\kappa_c^* > \kappa_c$. 
Using the three-loop $\beta$-function for the SR case~\cite{Foakes1987ThreeLoop,Foakes1988ThreeLoop,Amit2005FieldTheory},
we have ($g=1/K$)
\begin{equation}
\beta_{\rm RG} (g) = b g^2 + c g^3 + d g^4 + \mathcal{O}(g^5),
\label{eq:beta_RG}
\end{equation}
with
$b = \kappa_c - \kappa$, $c=\kappa_c/2\pi$ and $d=c (N+2)/8\pi$.
For $\kappa > \kappa_c$, the $\beta$-function admits an unstable critical fixed point $g^* > 0$ satisfying $\beta_{\rm RG}(g^*) = 0$. Solving for $g^*$ perturbatively for small $|b|/c$, we obtain
\( g^* = -\frac{b}{c} - \frac{d b^2}{c^3} + \mathcal{O}(b^3) \),
and, from the linearized RG flow around $g^*$, we obtain $\nu  \simeq  (c/b^2) $. For $N=3$, this gives
\begin{equation}
1/\nu \approx 4\pi^2 (\kappa_c^* - \kappa_c)^2 ,
\end{equation}
which indicates a very weak transition, consistent with the numerical estimate $1/\nu = 0.13(2)$.

In the low-\(T\) phase (\(\kappa>\kappa_c^*\)), the LR enhancement dominates over the SR
softening, making the spin-wave analysis reliable.
The low-energy physics of the \(O(3)\) NLSM is governed by its transverse modes, and the LR interaction modifies the kernel of
each component in the same way, giving the infrared kernel
\( \gamma(k)\simeq \kappa \,k^{2}\ln(1/k)\), with $\kappa$ replaced by $\kappa_{\rm 1L}=\kappa-\kappa_c>0$ at the one-loop level.
The two components decouple at the Gaussian level, and the softening induced
by their mutual interactions is absorbed into an effective coupling. 
Thus, we can treat them independently and consider the standard Gaussian spin-wave theory 
for a single real phase field~\cite{Pan2026log-QLRO}. 
With $\kappa_R$ denoting the renormalized spin stiffness beyond the one-loop level,
the spin-spin correlation function decays as
\begin{equation}
C(r)\sim(\ln r)^{-\eta_\ell},\qquad
\eta_\ell=(N-1)/(2\pi\kappa_R),
\end{equation}
which is slower than any power law but still vanishes in the thermodynamic limit. The low-\(T\) phase exhibits a log-QLRO rather than a true LRO. As \(L\) increases, the
squared magnetization decays as
\(\langle M^2\rangle\sim(\ln L)^{-\eta_\ell}\), and its Fourier counterpart
scales as \(\langle M_k^2\rangle\sim(\ln L)^{-\eta_\ell-1}\), 
while the correlation-length ratio grows logarithmically as 
\((\xi/L)^2\simeq s \ln L\), with $s=\kappa_R/(N-1)$ (see SM for derivations).
This leads to an identity 
\begin{equation}
{\cal P} \equiv 2\pi\kappa_{\rm R}\eta_\ell=(N-1),
\end{equation}
where $\kappa_{\rm R} \propto 1/T$ while $\eta_\ell \propto T$ for $T\to 0$.

{\it Numerical evidence for log-QLRO} ---
We further verify the log-QLRO phase from two distinctive finite-size predictions: \(\langle M^2\rangle \sim (\ln L)^{-\eta_\ell}\) and \((\xi/L)^2 \simeq s\ln L\).
Figure~\ref{fig:logqlro}(a) shows the log-log plot of \(\langle M^2\rangle\) versus \(\ln(L/L_0)\). 
The data follow approximately straight lines, and the estimated values of $\eta_\ell$ are given in Table~\ref{tab:logqlro}. 
The values at \(\beta=4\) and 8 are consistent with the linear scaling \(\eta_\ell\propto T\), while corrections are visible at higher temperature (\(\beta=2\)).
Figure~\ref{fig:logqlro}(b) plots \((\xi/L)^2\) against \(L\) on a semi-log scale. 
The data exhibit excellent linearity, and the slopes are well described by the linear relation $s=a+b \beta$ [inset of Fig.~\ref{fig:logqlro}(b)].
Table~\ref{tab:logqlro} summarizes the fitted values of \(\eta_\ell\) and \(\kappa_{\rm R}=s(N-1)\), and their product \(\mathcal{P}=2\pi \kappa_R \eta_\ell\).
For each of \(\beta=2,4,8\), the product is consistent with the predicted identity \(\mathcal{P}=2\).

\begin{figure}[t]
    \centering
    \includegraphics[width=\linewidth]{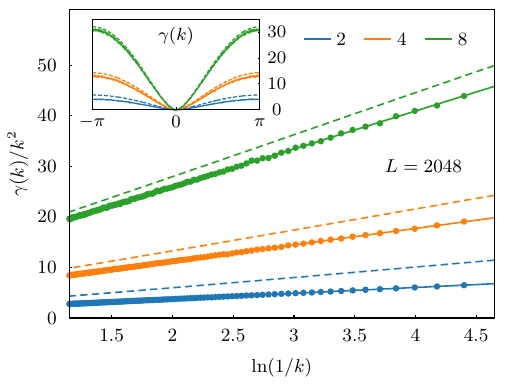}
    \vspace{-20pt}
    \caption{Effective spin-wave stiffness kernel of the 2D LR-Heisenberg model at \(\sigma=2\). Main panel: \(\gamma(k)/k^2\) versus \(\ln(1/k)\) for \(L=2048\) and \(\beta=2,4,8\), with \(\gamma(k)=(N-1)\langle M^2\rangle/\chi(k)\). Symbols are simulation data, solid lines are linear fits, and dashed lines show the one-loop reference \(\kappa-\kappa_c\) with the bare \(\kappa\) computed from the LR interactions. Inset: \(\gamma(k)\) over the full momentum range \(k\in[-\pi,\pi]\).}
    \label{fig:kernel}
    \vspace{-10pt}
\end{figure}

Because \(M\equiv\sqrt{\langle M^2\rangle}\) decreases only algebraically in \(\ln L\), it remains virtually finite at all realistic scales.
At \(\beta=2\), reducing \(M\) from \(0.70\) at \(L=4096\) to \(0.32\) requires \(L\sim10^{380}\) [Fig.~\ref{fig:logqlro}(a)]; at \(\beta=4\), reducing \(M\) from \(0.87\) to \(0.3\) requires \(L\sim e^{10^9}\).
Even for \(N_A\sim10^{23}\) spins, \(M\approx0.81\) at \(\beta=4\), only about a \(7\%\) reduction from \(L=4096\).
Thus, without spin-wave guidance, finite-size measurements inevitably mimic true LRO: the extended Mermin--Wagner theorem is not violated, yet its consequences are unobservable in practice.

Beyond the finite-size signatures, the infrared kernel is directly accessible in simulations. 
We extract \(\gamma(k)=(N-1) \langle M^{2}\rangle/\chi(k)\) from the momentum-dependent susceptibility \(\chi(k)\) over the full lattice axis, unlike \(\xi\), which uses only \(\chi_{k}=\chi(k_{\min})\). Figure~\ref{fig:kernel} shows
\(\gamma(k)/k^{2}\) versus \(\ln(1/k)\) for \(L=2048\) and
\(\beta=2,4,8\); the solid lines are linear fits confirming
\(\gamma(k)/k^{2}\simeq \kappa_R \ln(1/k)\), as detailed in the SM. The dashed
lines show the one-loop reference \(\kappa_{\rm 1L}=\kappa-\kappa_{c}\),
with \(\kappa=(\pi/2)\beta b\), \(b\simeq0.6637\), and
\(\kappa_{c}=1/(2\pi)\). Here \(b\) is the all-site
normalization of \(f(r)=b/r^{4}\) fixed by
\(4=\sum_{j\neq0}f(r_{0j})\) at \(\sigma=2\).
The relative deviation of \(\kappa_{\mathrm{R}}\) in Table~\ref{tab:logqlro} from
\(\kappa-\kappa_{c}\) decreases from about \(53\%\) at \(\beta=2\)
to \(25\%\) at \(\beta=4\) and \(11\%\) at \(\beta=8\), showing that the
spin-wave analysis becomes more reliable at lower temperature.

{\it Conclusion and outlook} --- 
Using large-scale simulations, an adiabatic RG analysis, and a spin-wave treatment with a logarithmic kernel, we have shown that the 2D long-range (LR) Heisenberg model at \(\sigma=2\) undergoes a finite-\(T\) continuous transition into a logarithmic quasi-long-range ordered (log-QLRO) phase, driven purely by non-topological spin-wave fluctuations. This transition transcends the Ginzburg-Landau-Wilson paradigm: the low-\(T\) phase does not admit a nonzero local order parameter in the thermodynamic limit, and the inverse correlation-length exponent \(1/\nu \propto (\kappa_c^*-\kappa_c)^{2}\) is non-universal.

In the low-\(T\) log-QLRO phase, the magnetization vanishes only in the strict thermodynamic limit, while \(C(r)\sim(\ln r)^{-\eta_\ell}\) decays so slowly that the phase is practically indistinguishable from true LRO at finite sizes. The distinctive scalings \((\xi/L)^2\simeq [\kappa_{\rm R}/(N-1)]\ln L\) and \(\langle M^2\rangle\sim(\ln L)^{-\eta_\ell}\) test the spin-wave relation \(2\pi \kappa_{\rm R}\eta_\ell=N-1\) and distinguish log-QLRO from true LRO; their quantitative consistency with the numerical data provides mutual validation of the two approaches.

A close parallel is the extraordinary-log boundary criticality of 3D \(O(N)\) models~\cite{Metlitski2022,Hu2021,ParisenToldin2021}, where a \(k^{2}\ln(1/k)\) boundary kernel is induced indirectly by fine-tuning the 3D bulk to criticality. 
Here the same kernel arises directly from LR interactions and its strength grows with \(\beta\), 
so it is tunable with temperature. This produces not only a low-\(T\)  log-QLRO phase but also a continuous transition into it, with a non-universal correlation-length exponent \(\nu\). Such interactions are accessible in quantum simulators and Rydberg arrays~\cite{Monroe2013,Browaeys2023}, where $(\xi/L)^{2}\simeq [\kappa_{\rm R}/(N-1)] \ln L$ provides a direct finite-size diagnostic.

Future directions include: (i) designing Hamiltonians with tunable SR/LR competition to test the non-universality of \(\nu\); (ii) a functional RG treatment in which \(\kappa\) is allowed to flow; and (iii) extensions to quantum long-range models, such as the 1D LR Bose-Hubbard model~\cite{Gupta2026BoseHubbard}, and to experimental platforms such as Rydberg atom arrays, where the \((\xi/L)^2\) diagnostic may prove particularly valuable.

\begin{acknowledgments}
{\it Acknowledgments} --
We thank Xianzhi Pan, Luis Walther, Zhiyi Li and Kun Chen for helpful discussions.
We acknowledge support from the National Natural Science Foundation of China (NSFC) under Grant No. 12204173 and No. 12275263, as well as Quantum Science and Technology-National Science and Technology Major Project (under Grant No. 2021ZD0301900). YD is also supported by the Natural Science Foundation of Fujian Province of China (Grant No. 2023J02032). ZF is also supported by the National Natural Science Foundation of China (NSFC) under Grant No. 12504265. L.P. acknowledges support from the Deutsche Forschungsgemeinschaft (DFG, German Research Foundation) under Germany’s Excellence Strategy – EXC-2111 – project number 390814868.
\end{acknowledgments}

\appendix
\section*{End Matter: Reliability of the Adiabatic RG and Spin-Wave Analysis}

Here we argue that the adiabatic RG and the  spin-wave expansion are asymptotically controlled in the low-$T$ 
log-QLRO phase, and identify the terms that limit their quantitative use near \(T_c\).

\paragraph{Enhancement versus softening.}
Within the adiabatic approximation, the running stiffness satisfies Eq.~(\ref{eq:Kflow}) at the one-loop level.
The LR enhancement is \(\kappa\propto\beta\), while the SR softening \(\kappa_c 
=(N-2)/2\pi \) is a $T$-independent pure number. 
At low temperature \(\kappa\gg\kappa_c\), so \(K\) grows logarithmically in the infrared. 

The three-loop beta function is given by Eq.~(\ref{eq:beta_RG}), which, 
in terms of \(K\), reads
\begin{equation}
\frac{dK}{d\ell}
=(\kappa-\kappa_c)-\frac{c}{K}-\frac{d}{K^2}+\mathcal{O}(K^{-3}).
\end{equation}
The higher-loop softening terms are suppressed by powers of \(1/K \propto T \). 
For \(N=3\) there are no stable vortices, so these terms arise purely from anharmonic spin-wave interactions.

\paragraph{Renormalization of \(\kappa\).}
Beyond the adiabatic approximation, higher-order vertices generate a flow of \(\kappa\) itself. Normalizing the transverse field by \(\sqrt{\kappa}\), the effective quartic coupling is
$g_4^{\mathrm{eff}}\sim 1/\kappa \propto T$.
A one-loop power-counting estimate gives
\begin{equation}
\frac{\delta\kappa}{\kappa}
\sim (g_4^{\mathrm{eff}})^2\ln\ln L
\sim \frac{1}{\kappa^2}\ln \ell 
\propto T^2\ln \ell,
\end{equation}
where the double logarithm reflects the \(\ln\ln r\) propagator of the log-QLRO phase and \(\ell=\ln L\) is the logarithmic RG scale. Differentiating with respect to \(\ell\) yields
\begin{equation}
\frac{1}{\kappa}\frac{d\kappa}{d\ell}
 \sim \frac{1}{\kappa^2}\frac{d(\ln\ell)}{d\ell}
= \frac{1}{\kappa^2\ell}
\propto \frac{T^2}{\ell}.
\end{equation}
Thus the renormalization of \(\kappa\) becomes progressively weaker as temperature decreases, and the Monte Carlo results approach the theoretical prediction \(\kappa_R\approx\kappa -\kappa_c \propto\beta\) at low temperature, as seen in Table~\ref{tab:logqlro}.

The suppression of the higher-loop softening and of the \(\kappa\) renormalization share the same origin: both are controlled by the small nonlinear coupling \(1/K\sim1/\kappa\propto T\). The former is the feedback of this coupling on the spin stiffness; the latter is its effect on the coefficient of the logarithmic kernel. 

\paragraph{Range of validity.}
In the low-\(T\) log-QLRO phase, higher-order corrections vanish as
\(T\to0\), so the adiabatic RG and the Gaussian spin-wave analysis
are asymptotically controlled. 
In the critical region, \(\kappa\) is renormalized by the same fluctuations that soften the
stiffness. 

An illustration is provided by the estimate of $\beta_c$.
The one-loop condition \(\kappa=\kappa_{c}=(N-2)/(2\pi)\), together with
\(\kappa=(\pi/2)\beta b\), where \(b\simeq0.6637\) is fixed by
\(\sum_{j\neq0}f(r_{0j})=4\), gives for \(N=3\)
\begin{equation}
    \beta_{c}^{\rm 1L}
    =\frac{2\kappa_{c}}{\pi b}
    =\frac{1}{\pi^{2}b}
    \simeq 0.152,
    \qquad
    T_{c}^{\rm 1L}\simeq 6.6 , 
\end{equation}
which is more than eight times the Monte Carlo value \(T_{c}\simeq0.78\)
[\(\beta_{c}=1.29(2)\)]. 
For asymptotically large $N$, the critical condition also predicts
\begin{equation}
    T_{c}\propto 1/N ,
\end{equation}
which reduces to the exact result for the spherical model for $N \to \infty$.

In short, the spin-wave and RG analyses are
qualitatively reliable throughout the whole \(T\le T_{c}\) region,
and their quantitative predictions improve progressively as \(T\)
decreases.  

\bibliography{ref_fix}

\end{document}


\definecolor{reviewpurple}{rgb}{0.5,0,0.5}

\clearpage
\onecolumngrid
\setcounter{equation}{0}
\setcounter{figure}{0}
\setcounter{table}{0}
\setcounter{page}{1}
\renewcommand{\thetable}{S\Roman{table}}
\renewcommand{\theequation}{S\arabic{equation}}
\renewcommand{\thefigure}{S\arabic{figure}}
\renewcommand{\theHtable}{S\Roman{table}}
\renewcommand{\theHequation}{S\arabic{equation}}
\renewcommand{\theHfigure}{S\arabic{figure}}

\begin{center}
{\large \bf Supplemental Material for\\[0.2em]
``Finite-Temperature Phase Transition and Logarithmic Quasi-Long-Range Order \\ 
in the Two-dimensional Long-range Heisenberg Model''}\\[1em]
\end{center}

\noindent
This Supplemental Material is organized as follows. Section I presents the
field-theoretic derivations of the finite-size scaling in the spin-wave
description and the log-QLRO phase, including the infrared kernel, the
transverse propagator, the spin-spin correlation function, and the
momentum-dependent susceptibilities, together with the numerical strategy
to test the spin-wave relation \(\mathcal{P}=2\pi\kappa_R\eta_\ell=N-1\).
Section II reports the Monte Carlo finite-size scaling analysis of the 2D
LR Heisenberg model, including the determination of critical points and
exponents, the verification of the log-QLRO scaling, and the
Goldstone-mode analysis for \(\sigma<2\).

\section{Finite-size scaling in spin-wave description and log-QLRO phase}

{\it Non-compact lattice field model and infrared kernel in spin-wave description}. 
We consider the classical $O(N)$ Heisenberg model on a two-dimensional
square lattice of linear size \(L\) with periodic boundary conditions, with reduced Hamiltonian
$\beta\mathcal{H} = -\beta \sum_{i<j} f(r_{ij}) \bm{S}_i \cdot \bm{S}_j$, where $f(r) = b/r^{2+\sigma}$.
For the low-temperature spin-wave expansion, we parameterize fluctuations about a uniformly aligned reference state as $\bm{S} = (\bm{\pi}, \sqrt{1-\bm{\pi}^2})$,
where $\bm{\pi} = (\pi_1, \dots, \pi_{N-1})$ represents the $N-1$ transverse spin-wave modes.
Expanding $\bm{S}_i \cdot \bm{S}_j \approx 1 - \frac{1}{2}\sum_{a=1}^{N-1}(\pi_{a,i}-\pi_{a,j})^2$
to quadratic order, the Hamiltonian decouples into $N-1$ independent non-compact Gaussian field modes
$\pi_a$, each governed by the reduced Hamiltonian
\begin{equation}
\mathcal{H}_a = \frac{\lambda}{2} \sum_{i<j} \frac{1}{r_{ij}^{2+\sigma}}
(\pi_{a,i} - \pi_{a,j})^2 ,
\label{eq:SM:H}
\end{equation}
where \(r_{ij}\) is the distance between sites \(i\) and \(j\) and the sum
runs over all pairs of sites, with coupling $\lambda = \beta b$. 
In the limit \(\sigma \to \infty\), the model
reduces to the standard short-range Gaussian model.

In the continuum limit, the effective action for each transverse mode in momentum space reads
\begin{equation*}
S_a = \frac{1}{2} \int \frac{d^2 k}{(2\pi)^2} \, \gamma(k) \, |\pi_a(k)|^2 ,
\end{equation*}
where the stiffness kernel is
\[
\gamma(k)
= \lambda \int d^2 r \, \frac{1 - \cos(\mathbf{k}\cdot \mathbf{r})}
{r^{2+\sigma}} .
\]
Using polar coordinates and the Bessel function \(J_0\), we have
\begin{equation}
\gamma(k)
= 2\pi \lambda \, k^{\sigma}
\int_{k a}^{\infty} \frac{du}{u^{1+\sigma}} [1 - J_0(u)] ,
\label{eq:SM:gammaLRint}
\end{equation}
where \(a\) is the ultraviolet lattice cutoff and $u = kr$.
For \(\sigma \ge 2\), the integral diverges at its lower limit
as \(ka \to 0\). We therefore use the small-\(u\) expansion
\(J_0(u) \simeq 1 - u^2/4\) to extract its leading dependence on \(ka\).
For \(0 < \sigma < 2\), the integral converges to a finite positive
constant as \(ka \to 0\).
Setting \(a = 1\) for convenience, we obtain the leading infrared behavior
\begin{equation}
\gamma(k) \simeq
\begin{cases}
B_\sigma \, k^2 , & \sigma > 2 , \\[4pt]
A_\sigma \, k^{\sigma} , & 0 < \sigma < 2 , \\[4pt]
\kappa \, k^2 \ln(1/k) , & \sigma = 2 ,
\end{cases}
\label{eq:SM:kernels}
\end{equation}
where \(B_\sigma\) and \(A_\sigma\) are positive constants, and \(\kappa = \pi \lambda/2 > 0  \).
For \(\sigma > 2\), the nonanalytic long-range contribution is subleading, while the leading kernel remains proportional to \(k^2\), with a stiffness \(B_\sigma\) that includes contributions from the long-range tail. Thus the system retains the standard short-range low-temperature behavior.
For \(0 <\sigma < 2\), the
nonanalytic term \(k^{\sigma}\) dominates in the infrared. At the marginal
case \(\sigma = 2\), the long-range interaction generates the logarithmic
kernel
\begin{equation}
\gamma(k) \simeq \kappa \, k^2 \ln(1/k) , \qquad \kappa = \pi \lambda/2 = \frac{\pi}{2}\beta b , 
\label{eq:SM:logkernel}
\end{equation}
which is the starting point for the finite-size analysis in the following
sections.

{\it Finite-size transverse correlation function in spin-wave description at $\sigma=2$}.
On a finite \(L\times L\) lattice with periodic boundary conditions, the single-component transverse propagator (with the uniform mode omitted and no summation over $a$) is defined as
\begin{equation}
G(r,L)
= \langle \pi_a(\mathbf{r}) \pi_a(0) \rangle
= \frac{1}{L^2} \sum_{\mathbf{k} \neq 0}
\frac{e^{i \mathbf{k}\cdot \mathbf{r}}}{\gamma_L(k)} ,
\label{eq:SM:Gdef}
\end{equation}
where the allowed momenta are
\(\mathbf{k} = (2\pi n_x/L, 2\pi n_y/L)\) with
\(n_x,n_y = -L/2+1,\dots,L/2\), and \(\gamma_L(k)\) is the finite-size
stiffness kernel that reduces to \(\gamma(k)\) for \(L \to \infty\). 
For \(L \gg 1\), we approximate the momentum sum by a continuum integral
with an infrared cutoff \(k_{\min} = 2\pi/L\) and an ultraviolet cutoff
\(\Lambda \sim 1\),
\begin{equation*}
G(r,L)
\simeq \frac{1}{(2\pi)^2} \int_{k_{\min}}^{\Lambda} d^2 k \,
\frac{e^{i \mathbf{k}\cdot \mathbf{r}}}{\gamma(k)} \simeq \frac{1}{2\pi \kappa}
\int_{k_{\min}}^{\Lambda} \frac{dk}{k} \,
\frac{J_0(kr)}{\ln(1/k)} ,
\end{equation*}
where we have performed the angular integral, \(\int_0^{2\pi} d\theta \, e^{i k r \cos\theta} = 2\pi J_0(kr)\).
The behavior of \(G(r,L)\) depends on the range of \(r\), and we analyze
the three regimes separately.

\begin{itemize}
\item Regime 1: \(r \in O(1)\).
For \(r\) of order the lattice spacing, the Bessel function satisfies \(J_0(kr) \simeq 1\) throughout the infrared region \(k \lesssim \Lambda \sim 1\). The leading infrared divergence comes entirely from the lower integration cutoff \(k_{\min} = 2\pi/L\),
\begin{equation}
G(r,L)
\simeq \frac{1}{2\pi \kappa}
\int_{k_{\min}}^{\Lambda} \frac{dk}{k \ln(1/k)} + \mathcal{O}(1)
= \frac{1}{2\pi \kappa} \ln\ln(1/k_{\min}) + \mathcal{O}(1) \simeq \frac{1}{2\pi \kappa} \ln\ln L + O(1) ,
\qquad r \in O(1) .
\label{eq:SM:Greg1final}
\end{equation}

\item Regime 2: \(1 \ll r \ll L\).
In this regime, \(k_{\min} \ll 1/r \ll \Lambda\), so the integral splits
into two parts:
\begin{equation*}
G(r,L)
\simeq \frac{1}{2\pi \kappa}
\left[
\int_{k_{\min}}^{1/r} \frac{dk}{k} \frac{J_0(kr)}{\ln(1/k)}
+ \int_{1/r}^{\Lambda} \frac{dk}{k} \frac{J_0(kr)}{\ln(1/k)}
\right] .
\end{equation*}
For \(k < 1/r\), \(J_0(kr) \simeq 1 - (kr)^2/4\),
with the quadratic term giving only an \(O(1/\ln r)\)
correction. The leading contribution is
\begin{equation*}
\int_{k_{\min}}^{1/r} \frac{dk}{k \ln(1/k)}
= \ln\ln(1/k_{\min}) - \ln\ln r
\simeq \ln\ln L - \ln\ln r .
\end{equation*}
For \(k > 1/r\), \(J_0(kr)\)
oscillates and decays, contributing another \(O(1/\ln r)\)
correction.  Hence
\begin{equation}
G(r,L)
\simeq \frac{1}{2\pi \kappa}
\left( \ln\ln L - \ln\ln r \right) + O(1/\ln r) ,
\qquad 1 \ll r \ll L .
\label{eq:SM:Greg2}
\end{equation}
Thus \(G(r,L)\) decreases with increasing \(r\), with a logarithmic
correction \(\ln\ln r\), while retaining the
infrared divergence \(\ln\ln L\) from the finite system size.


\item Regime 3: \(r \mathrel{\sim} \alpha L\).
When \(r\) is comparable to \(L\), the continuum integral approximation
breaks down because \(k_{\min} r = 2\pi \alpha\) is of order unity. Choosing \(\mathbf{r}=(\alpha L, 0)\) along a principal lattice axis without loss of generality, we return to the discrete sum,
\begin{equation*}
G(\alpha L, L)
= \frac{1}{L^2} \sum_{\mathbf{k} \neq 0}
\frac{e^{i k_x \alpha L}}{\gamma_L(k)} .
\end{equation*}
The dominant contribution comes from the smallest momenta,
\(\mathbf{k}=(2\pi/L)\mathbf{n}\), where
\(\mathbf{n}=(n_x,n_y)\) and \(|\mathbf{n}|=O(1)\).
For these modes, we approximate
\(\gamma_L(k)\simeq
\kappa(2\pi/L)^2|\mathbf{n}|^2\ln L\). Hence
\begin{equation}
G(\alpha L, L)
\simeq \frac{1}{4\pi^2\kappa\ln L}
\sum_{\mathbf{n}\neq\mathbf{0}}
\frac{e^{i2\pi\alpha n_x}}{|\mathbf{n}|^2}
= O(1/\ln L) \to 0
\qquad (L \to \infty) .
\label{eq:SM:Greg3}
\end{equation}
Here the weighted Fourier sum, understood with the lattice cutoff,
provides a bounded prefactor at fixed \(0<\alpha<1\).
Therefore \(G(r,L)\) vanishes when \(r\) approaches the system size.

\end{itemize}

Collecting the three regimes, the finite-size transverse propagator at
\(\sigma=2\) behaves as
\begin{equation}
G(r,L) \simeq
\begin{cases}
\displaystyle
\frac{1}{2\pi \kappa} \ln\ln L + O(1) ,
& r \in O(1) , \\[8pt]
\displaystyle
\frac{1}{2\pi \kappa}
\left( \ln\ln L - \ln\ln r \right) + O(1/\ln r),
& 1 \ll r \ll L , \\[8pt]
\displaystyle
O(1/\ln L) \to 0 ,
& r = \alpha L .
\end{cases}
\label{eq:SM:Gsummary}
\end{equation}
This structure—an infrared divergence \(\ln\ln L\) that is gradually
canceled by the \(r\)-dependent term, \(-\ln\ln r\), and finally vanishes
at \(r \sim L\)—is the origin of the logarithmic quasi-long-range order (log-QLRO)
at \(\sigma=2\). The corresponding transverse fluctuations and their
consequences for the spin-spin correlation function \( C(r,L)\) are discussed below.

\textit{Finite-size spin-spin correlation function}. 
For the \(O(N)\) spin model, the spin vector is parameterized by the \(N-1\) transverse Goldstone modes \(\bm{\pi} = (\pi_1, \dots, \pi_{N-1})\) as \(\bm{S} = (\bm{\pi}, \sqrt{1-\bm{\pi}^2})\).
To second order in fluctuations, \(\bm{S}(\mathbf{r})\cdot\bm{S}(0) \approx 1 - \frac{1}{2}\sum_{a=1}^{N-1} [\pi_a(\mathbf{r}) - \pi_a(0)]^2\).
Exponentiating the Gaussian fluctuations, the spin-spin correlation function is
\begin{equation}
C(r,L)
= \langle \bm{S}(\mathbf{r})\cdot\bm{S}(0) \rangle
\simeq e^{-\frac12 \sum_{a=1}^{N-1} \langle [\pi_a(\mathbf{r}) - \pi_a(0)]^2 \rangle} .
\label{eq:SM:Cdef}
\end{equation}
At the Gaussian level, the \(N-1\) transverse components decouple and each has propagator \(G(r,L)\),
\begin{equation*}
\langle [\pi_a(\mathbf{r})-\pi_a(0)]^2 \rangle
= 2[G(0,L)-G(r,L)] ,
\end{equation*}
with the single-mode difference given, in the continuum approximation, by
\[
G(0,L)-G(r,L)
\simeq \frac{1}{2\pi\kappa}
\int_{k_{\min}}^\Lambda
\frac{1-J_0(kr)}{k\ln(1/k)}\,dk .
\]
Summing over all \(N-1\) modes, the spin-spin correlation function becomes
\begin{equation}
C(r,L)
\simeq e^{-(N-1)[G(0,L)-G(r,L)]} .
\label{eq:SM:CfromG}
\end{equation}

We now evaluate Eq.~\eqref{eq:SM:CfromG} in the three regimes of \(r\)
identified for \(G(r,L)\).

\begin{itemize}
\item Regime 1: \(r \in O(1)\).
For \(r\) of order the lattice spacing, \(1/r\sim\Lambda\), so the
integral for \(G(0,L)-G(r,L)\) above is dominated by the
ultraviolet region \(k\sim1/r\). The infrared divergence \(\ln\ln L\) appears identically
in \(G(0,L)\) and \(G(r,L)\), and cancels in the difference. Therefore
\begin{equation}
G(0,L) - G(r,L) = O(1) ,
\end{equation}
and hence
\begin{equation}
C(r,L) \simeq e^{-(N-1)[G(0,L) - G(r,L)]} = O(1) ,
\qquad r \in O(1) .
\label{eq:SM:Creg1}
\end{equation}
In particular, \(C(r,L) \to 1\) as \(r \to 0\), as required by the
normalization of the spin field. 

\item Regime 2: \(1 \ll r \ll L\).
In this regime,
\begin{equation*}
G(0,L) - G(r,L)
\simeq \frac{1}{2\pi \kappa} \ln\ln r ,
\end{equation*}
and therefore
\begin{equation}
C(r,L) \simeq (\ln r)^{-\eta_\ell} ,
\qquad
\eta_\ell = \frac{N-1}{2\pi \kappa} .
\label{eq:SM:Creg2}
\end{equation}
This is the thermodynamic-limit log-QLRO correlation function. Its decay
is slower than any power law, but it still vanishes as \(r \to \infty\).

\item Regime 3: \(r = \alpha L\).
When \(r\) is comparable to \(L\), the transverse propagator satisfies
\(G(\alpha L,L) \sim O(1/\ln L) \to 0\), while
\(G(0,L) \simeq \frac{1}{2\pi \kappa} \ln\ln L\). Hence
\begin{equation*}
G(0,L) - G(\alpha L,L)
\simeq \frac{1}{2\pi \kappa} \ln\ln L ,
\end{equation*}
and
\begin{equation}
C(\alpha L,L) \simeq (\ln L)^{-\eta_\ell} .
\label{eq:SM:Creg3}
\end{equation}
Thus, at separations of order \(L\), the correlation scales as a finite-size value consistent with the logarithmic decay in regime 2.
\end{itemize}

Collecting the three regimes,  the spin-spin correlation function for \(r\gg1\)
can be written in the scaling form
\begin{equation}
C(r,L)\simeq(\ln r)^{-\eta_\ell}\,
\mathcal{F}_C\!\left(\frac{r}{L}\right),
\qquad
\mathcal{F}_C(x\to0)\to1, \qquad \mbox{with} \, \eta_\ell=(N-1)/(2\pi\kappa) .
\end{equation}

\textit{Finite-size scaling of momentum-dependent spin susceptibility}.
The spin susceptibility at momentum \(\mathbf{k}\) is
\begin{equation}
\chi(k,L)
= \sum_{\mathbf{r}} e^{i \mathbf{k}\cdot \mathbf{r}} C(r,L)
\simeq e^{-(N-1)G(0,L)} \sum_{\mathbf{r}} e^{i \mathbf{k}\cdot \mathbf{r}}
e^{(N-1)G(r,L)} .
\label{eq:SM:chiDef}
\end{equation}
We expand \(e^{(N-1)G(r,L)}\) in powers of \(G(r,L)\),
\begin{equation}
e^{(N-1)G(r,L)}
= 1 + (N-1)G(r,L) + \frac{1}{2} (N-1)^2 G(r,L)^2 + \dots . 
\label{eq:SM:expG}
\end{equation}
Using the orthogonality relation
\begin{equation*}
\sum_{\mathbf{r}} e^{i \mathbf{k}\cdot \mathbf{r}}
= L^2 \delta_{\mathbf{k},0} ,
\end{equation*}
the constant term in Eq.~\eqref{eq:SM:expG} contributes only at
\(\mathbf{k}=0\), while for \(\mathbf{k}\neq0\) it vanishes. The linear
term gives
\begin{equation*}
(N-1)\sum_{\mathbf{r}} e^{i \mathbf{k}\cdot \mathbf{r}} G(r,L)
= \frac{N-1}{\gamma_L(k)} ,
\qquad \mathbf{k} \neq 0 ,
\end{equation*}
which reflects the mode multiplicity: each of the \(N-1\) transverse Goldstone modes contributes equally to the non-zero momentum susceptibility. Thus the leading contributions to \(\chi(0,L)\) and
\(\chi(k_{\min},L)\) come from different orders in the expansion.

For \(\mathbf{k}=0\), the constant term gives \(L^2\), while the linear
term vanishes because \(\sum_{\mathbf{r}} G(r,L)=0\) for \(\mathbf{k}=0\).
The first subleading contribution comes from the quadratic term,
\begin{equation*}
\frac{(N-1)^2}{2} \sum_{\mathbf{r}} G(r,L)^2
= \frac{(N-1)^2}{2L^2} \sum_{\mathbf{q} \neq 0}
\frac{1}{\gamma_L(q)^2} .
\end{equation*}
For \(\sigma=2\), the infrared kernel is
\(\gamma_L(q) \simeq \kappa q^2 \ln(1/q)\), and the sum is dominated by
the infrared region \(q \sim 2\pi/L\). In the thermodynamic limit, it leads to
\begin{equation*}
\frac{(N-1)^2}{2L^2} \sum_{\mathbf{q} \neq 0}
\frac{1}{\gamma_L(q)^2}
\sim \frac{L^2}{(\ln L)^2} ,
\end{equation*}
so that the relative correction to the constant term is of order \(1/(\ln L)^2\).
From the spin-wave result for $G(0,L)$, we have 
\begin{equation}
\chi(0,L)  \simeq L^2 (\ln L)^{-\eta_\ell}, \qquad \eta_\ell = (N-1)/(2\pi \kappa), 
\label{eq:SM:chi0}
\end{equation}
and the squared magnetization density \(\langle M^2 \rangle \equiv  \chi(0,L)/L^2  \) scales as
\begin{equation*}
\langle M^2 \rangle \simeq (\ln L)^{-\eta_\ell} .
\end{equation*}

For low-momentum modes with \(\mathbf{k} = 2\pi\mathbf{n}/L\) (fixed non-zero \(\mathbf{n}\)), the constant term is absent, and the linear term dominates:
\begin{equation}
\chi(k,L)
\simeq (N-1) e^{-(N-1)G(0,L)} \frac{1}{\gamma_L(k)} 
\simeq (N-1)\frac{\langle M^2\rangle}{\gamma_L(k)} ,
\qquad \mathbf{k} \neq 0 .
\label{eq:SM:chik}
\end{equation}
The quadratic term gives a relative correction of order \(1/\ln L\)
for the small-momentum modes considered here, which is subleading. At the smallest nonzero momentum, we have 
\begin{equation}
\chi(k_{\min},L)
\simeq \frac{N-1}{4\pi^2\kappa} \frac{L^2\langle M^2\rangle}{\ln L} \sim L^2 (\ln L)^{-\eta_\ell-1} ,
\label{eq:SM:chikfinal}
\end{equation}
and the squared Fourier magnetization density is
\begin{equation*}
\langle M_k^2 \rangle \sim (\ln L)^{-1-\eta_\ell} .
\end{equation*}

\textit{Finite-size scaling of correlation-length ratio \((\xi/L)^2\)}.
The two susceptibilities \(\chi(0,L)\) and \(\chi(k_{\min},L)\) share the
same global amplitude \(e^{-(N-1)G(0,L)}\), which therefore cancels in
their ratio:
\begin{equation}
\chi(0,L)/\chi(k_{\min},L) \simeq \frac{L^2 \gamma_L(k_{\min})}{N-1} \simeq \frac{\kappa}{N-1} (2\pi)^2 \ln L ,
\label{eq:SM:ratio1}
\end{equation}
with $k_{\min}=2\pi/L$.

The second-moment correlation length is defined as
\begin{equation}
\xi
= \frac{1}{2\sin(\pi/L)}
\sqrt{ \frac{\chi(0,L)}{\chi(k_{\min},L)} - 1 } .
\label{eq:SM:xidef}
\end{equation}
For large \(L\), \(2\sin(\pi/L) \simeq 2\pi/L\), so that
\begin{equation}
\left(\xi/L \right)^2 \simeq \frac{\kappa}{N-1} \ln L .
\label{eq:SM:xiLfinal}
\end{equation}

\textit{Numerical detection of log-QLRO}.
The Gaussian spin-wave analysis treats the transverse modes as independent free fields governed by the bare long-range coupling \(\kappa\), neglecting non-Gaussian fluctuations. For
the XY model, these non-Gaussian effects include vortex-antivortex pairs and higher-order spin-wave interactions; for \(O(N)\)
models with \(N>2\), they arise from higher-order spin-wave interactions, including couplings between different spin-wave components. In the low-temperature phase, although
such excitations are irrelevant in the renormalization-group sense, they
renormalize the bare coupling into an effective stiffness,
\begin{equation}
\kappa \to \kappa_R ,
\label{eq:SM:kappaR}
\end{equation}
without changing the logarithmic form of the infrared kernel,
\begin{equation}
\gamma_L(k) \simeq \kappa_R \, k^2 \ln(1/k) .
\label{eq:SM:kernelR}
\end{equation}
The log-QLRO phase therefore survives, but its parameters are
renormalized. This suggests a direct numerical test of the spin-wave
picture that does not rely on the bare value of \(\kappa\).

In particular, the stiffness kernel \(\gamma_L(k)\) and the renormalized stiffness \(\kappa_R\) extracted from Monte Carlo data already incorporate all non-Gaussian fluctuation effects. Inverting Eq.~\eqref{eq:SM:chik} gives the formula used in the main text to extract the stiffness kernel directly from numerical data:
\begin{equation}
\gamma_L(k) \simeq (N-1)\frac{\langle M^2\rangle}{\chi(k,L)} .
\label{eq:SM:gammaknum}
\end{equation}
In simulations, the susceptibility \(\chi(\mathbf{k},L)\) is evaluated along the principal lattice axes \(\mathbf{k}=(k,0)\) and \((0,k)\), and averaged over the four symmetry-equivalent directions to reduce statistical fluctuations.
On a system size \(L=2048\), the stiffness kernel is extracted directly via Eq.~\eqref{eq:SM:gammaknum}.
To extract the renormalized stiffness \(\kappa_R\) shown in Fig.~4 of the main text, the ratio \(\gamma_L(k)/k^2\) is fitted to \(a + \kappa_R \ln(1/k)\) within the infrared window \(0.04 \le k \le 0.16\). Accounting for both statistical errors and systematic uncertainties arising from the choice of fitting window, finite-momentum lattice corrections, and small-momentum finite-size corrections, we obtain \(\kappa_R = 1.15(2)\), \(3.3(1)\), and \(7.5(2)\) for \(\beta=2, 4, 8\), respectively. These momentum-space estimates agree well with the finite-size scaling estimates from \((\xi/L)^2\) in Table~II of the main text at low temperatures, while the modest difference at \(\beta=2\) may reflect remaining finite-size effects and additional temperature-dependent corrections in \(\langle M^2\rangle\).

In the presence of fluctuation renormalization, the parameters of the log-QLRO scaling are governed by the renormalized stiffness \(\kappa_R\):
\begin{equation}
\eta_\ell = \frac{N-1}{2\pi \kappa_R} , \qquad s = \frac{\kappa_R}{N-1} ,
\label{eq:SM:renorm_params}
\end{equation}
where \((\xi/L)^2 \simeq s\ln L\). At one-loop order in the continuum NLSM, \(\kappa_R\) is approximated by \(\kappa_{\rm 1L} = \kappa - \kappa_c\) with \(\kappa_c = 1/(2\pi)\) for \(N=3\), as discussed in the main text.
From the finite-size scaling of the squared magnetization, \(\langle M^2 \rangle \sim (\ln L)^{-\eta_\ell}\),
we extract the exponent \(\eta_\ell\). Independently, from the correlation-length ratio,
\( \left( \xi/L \right)^2 \simeq s \ln L  \),
we extract the slope \(s\), which determines the renormalized coupling \(\kappa_R = (N-1)s\). 
The spin-wave theory then predicts the universal relation
\begin{equation}
\mathcal{P} \equiv 2\pi \kappa_R \eta_\ell = 2\pi (N-1) s \eta_\ell = N-1 .
\label{eq:SM:Pnum}
\end{equation}
For the Heisenberg model ($N=3$), this predicts $\mathcal{P}=N-1=2$.
Verifying this relation in Monte Carlo data provides a direct test of the
log-QLRO scenario.

This test is particularly useful because it is insensitive to the bare
value of \(\kappa\) and to the detailed nature of the irrelevant
excitations. In the XY model, vortices are strongly suppressed by the
\((\ln L)^2\) energy cost, and together with higher-order spin-wave interactions they still contribute to the renormalization
of \(\kappa\). In the Heisenberg model, which does not support topologically stable vortices, higher-order spin-wave interactions play the analogous role. In both
cases, the infrared logarithmic kernel is preserved, and the relation
\(\mathcal{P}=N-1\) should hold if the spin-wave description is
self-consistent.

\section{Finite-size scaling analysis of the 2D LR Heisenberg model}
\label{sm_sec1}

In this section, we present the finite-size scaling (FSS) analysis of the 2D long-range Heisenberg model for $\sigma \le 2$. We first determine the critical properties, including the critical point and critical exponents, and then show the finite-size scaling analysis of the log-QLRO phase.

\paragraph{Critical point and critical exponents.}
We first present a finite-size scaling analysis of the critical properties of the 2D LR Heisenberg model for $\sigma \le 2$ (specifically, $\sigma=1.75$, $1.875$, and $2$). 

\begin{figure}[b]
    \centering
    \includegraphics[width=0.9\linewidth]{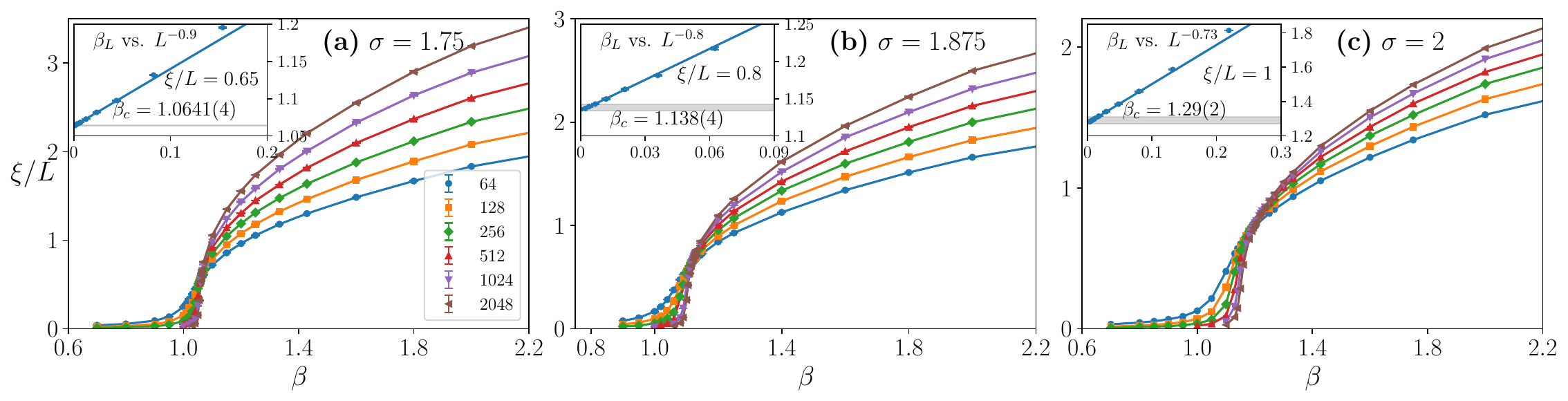}
    \caption{Demonstration of the second-order phase transition for $\sigma=1.75$ (a), $1.875$ (b) and $2$ (c). Main plots display the second-moment correlation-length ratio $\xi/L$ as a function of the inverse temperature $\beta$. The clear crossing behavior among curves with different $L$ signifies the existence of a second-order phase transition. The insets show the pseudo-critical inverse temperature $\beta_L$ when $\xi/L$ is fixed at $0.65, 0.8$ and 1 for $\sigma=1.75,1.875$ and $2$. In all cases, $\beta_L$ exhibits power-law convergence to a finite critical point.}
    \label{LRO3_critical}
\end{figure}

Figure~\ref{LRO3_critical} displays the critical characteristics for $\sigma = 1.75$, $1.875$, and $2$. The main panels show $\xi/L$ as a function of the inverse temperature $\beta$. In all cases, the pronounced crossing of curves for different system sizes $L$ signifies a second-order transition. By performing an FSS analysis near the crossing point, we can estimate the critical point $\beta_c$. Specifically, near criticality, the correlation-length ratio $Q \equiv \xi/L$—as a function of the distance to the critical point $\epsilon=\beta-\beta_c$, system size $L$, and the irrelevant scaling field $u$—admits a Taylor expansion of the form~\cite{Yao2025LRXY}:
\begin{equation}
    Q(\epsilon,u,L)=  a_0+\sum_{i=1}^{m}a_i(\epsilon L^{y_t})^i+b_1 L^{-\omega}
    + c_1\epsilon L^{y_t-\omega} + c_2\epsilon^2L^{2y_t-\omega},
    \label{CriticalPoint_eq}
\end{equation}
where the value of $m$ depends on the specific fitting process and is typically taken as $2$ or $3$; $y_t$ is the thermal scaling exponent; the $L^{-\omega}$ term comes from the irrelevant field $u$. Fitting $\xi/L$ to the equation above yields the estimate of the critical point. During the fitting procedure, we vary the fixed value of $\omega$ to explore possible fitting outcomes. For a given fitting window, we systematically adjust the minimum system size $L_\text{min}$ included in the fit. When the fitted parameters remain stable under these adjustments, and the reduced chi-square $\chi^2/\text{DF}$ (DF denotes the degrees of freedom) remains close to $1$, we consider the fit a plausible estimate of the critical point.

Table~\ref{CriticalPointFit} presents the detailed fitting results for $\sigma=1.75,1.875,2$. For each $\sigma$, we obtain two sets of plausible fits using different fixed values of $\omega$. By combining the results from both fits, we obtain a final estimate of the critical point and its associated uncertainty. Table~\ref {Kc_values} lists the estimated critical points for all parameter sets included in the phase diagram (Fig. 1 in the main text).

The insets of Fig.~\ref{LRO3_critical} illustrate the FSS of the pseudo-critical inverse temperature $\beta_L$, determined by the $\beta$ value at which $\xi/L$ is fixed near its critical value. For each $\sigma$, as $L$ increases, $\beta_L$ converges toward the thermodynamic critical point with a power-law decay $\beta_L - \beta_c\sim L^{- \omega_{\rm shift}}$, further confirming a finite-$T$ phase transition. Here, $\omega_{\rm shift}$ is the drift exponent describing the convergence of the pseudo-critical point, which is distinct from the correction-to-scaling exponent $\omega$ of the irrelevant field. Moreover, the asymptotic value to which $\beta_L$ converges is fully consistent with the critical-point estimates $\beta_c$ obtained from independent FSS analyses for each $\sigma$.

After locating the critical point, we proceed to determine the critical exponents for $\sigma=1.75, 1.875$, and $2$ using FSS at criticality. Given the scaling relations among critical exponents, only two are independent; here we focus on the anomalous dimension $\eta$ and the thermal exponent $y_t=1/\nu$, where $\nu$ denotes the correlation-length exponent. 

\begin{table}[h]
    \centering
    \caption{The fitting details for determining the critical point using Eq.~\eqref{CriticalPoint_eq}}
    \begin{tabular}{lllllllllllll}
    \hline\hline
        $~\sigma$ & $L_\mathrm{min}$ & $~~~~\beta_c$ & $~~y_t$ & $~~a_0$ & $~~a_1$ & $~~a_2$ & $~~a_3$ & $~~~b_1$ & $~~c_1$ & $~~c_2$ & $~\omega$ &  $\chi^2/\mathrm{DF}$  \\ \hline
        1.75 & 128 & 1.06452(9) & 0.29(1) & 0.773(4) & 3.3(3) & -26(8) & 187(49) & -0.354(7) & -2.9(5) & 67(21) & 0.15 & 25.3/29 \\
        & 256 & 1.0645(1) & 0.29(1) & 0.771(6) & 3.1(4) & -30(11) & 198(67) & -0.35(1) & -2.4(6) & 80(30) & & 16.4/21 \\
        &128 & 1.06374(5) & 0.34(1) & 0.678(1) & 1.9(1) & -7(2) & 67(18) & -0.355(7) & -1.9(4) & 36(11) & 0.3 & 28.2/29 \\ 
        &256 & 1.06385(7) & 0.34(1) & 0.683(2) & 1.8(2) & -8(2) & 75(22) & -0.38(1) & -1.5(5) & 47(16) & & 15.6/21 \\ \hline
        1.875 & 128 & 1.140(1) & 0.14(1) & 0.889(8) & 2.4(2) & -7(1) & - & -0.65(2) & -3.2(8) & - & 0.3&19.1/21 \\
        &256 & 1.141(1) & 0.13(2) & 0.89(1) & 2.6(4) & -9(2) & - & -0.68(4) & -4(1) & - & &14.3/16 \\
        &128 & 1.1349(5) & 0.16(1) & 0.823(4) & 2.1(2) & -5(1) & - & -0.74(1) & -3(1) & - & 0.4 &20.2/21 \\ 
        &256 & 1.1360(8) & 0.15(1) & 0.833(7) & 2.3(3) & -6(1) & - & -0.79(3) & -4(1) & - &&14.4/16 \\ \hline
        2 & 128 & 1.304(6) & 0.22(1) & 1.21(2) & 0.9(1) & 1.6(4) & 0.9(3) & -0.96(6) & -1.8(5) & -7(1) & 0.25 & 40.1/36 \\
        & 256 & 1.298(7) & 0.21(1) & 1.18(3) & 0.9(2) & 1.7(6) & 1.1(6) & -0.9(1) & -1.7(7) & -8(1) & &  33.2/30 \\
        &128 & 1.280(3) & 0.29(1) & 1.06(1) & 0.32(6) & 0.4(1) & 0.15(6) & -0.82(4) & -0.1(2) & -4.1(7) & 0.33 & 41.3/36 \\
        &256 & 1.279(5) & 0.29(1) & 1.05(1) & 0.30(8) & 0.4(1) & 0.16(8) & -0.75(8) & ~0.1(3) & -4.2(9) & & 33.3/30 \\ \hline
    \end{tabular}
    \label{CriticalPointFit}
\end{table}

\begin{table}[h]
    \centering
    \caption{\looseness=-1 Estimates of the critical inverse temperature $\beta_c$ for various $\sigma$ shown in Fig. 1 in the main text. Considering the normalization procedure $\sum_{j\neq 0} b/r_{0j}^{2+\sigma} = 4$, when $\sigma\to-2$, the critical point converges to its value in the complete graph: $\beta_c\to\frac34$.}
    \begin{threeparttable}
    \begin{tabular}{l|l||l|l}
    \hline\hline
        ~~$\sigma$ & ~~~~$\beta_c$ & ~~$\sigma$ & ~~~~$\beta_c$ \\ \hline
        0.25  & 0.758(1)      & 1.25     & 0.910(2)       \\
        0.45  & 0.768(1)     & 1.5   & 0.982(3)         \\
        0.65  & 0.790(3)      & 1.75   & 1.0641(4)       \\
        0.85  & 0.825(2)      & 1.875   & 1.138(4)       \\
        1     & 0.850(2)     & 2    & 1.29(2)         \\
     \hline\hline
    \end{tabular}
    \label{Kc_values}
    \end{threeparttable}
\end{table}

The standard procedure for extracting critical exponents typically involves two steps: first determining the critical point $\beta_c$ with high precision, and then performing an FSS analysis of appropriate observables at $\beta_c$ to obtain the exponents from their scaling behavior. It should be noted that although fitting Eq.~\eqref{CriticalPoint_eq} to $\xi/L$ data also yields an estimate of the thermal exponent $y_t$, such estimates should be treated with caution, particularly when $y_t$ is small. This is because the fitting form in Eq.~\eqref{CriticalPoint_eq} is highly nonlinear in the term $L^{y_t}$. Moreover, the estimated value of $y_t$ is strongly sensitive to the fitting choices, including the order $m$ of the expansion used, as well as whether correction terms -- such as $c_1 \epsilon L^{y_t-\omega}$, $c_2 \epsilon^2 L^{2y_t-\omega}$, etc. -- are included in the fitting function. 

In order to extract $\eta$ and $y_t$ more reliably, following the approach in Ref.~\cite{Yao2025LRXY}, we compute and fit two quantities: the susceptibility $\chi=L^2\langle M^2\rangle$ and the scaled energy-magnetization covariance $C_{em}$ between $M^2$ and the nearest-neighbor (NN) energy density $\varepsilon = - \frac{J_{\text{nn}}}{L^d} \sum_{\langle i,j \rangle} \bm{S}_i \cdot \bm{S}_j$, where $\langle i,j \rangle$ denotes NN pairs of spins and $J_{\text{nn}}$ is the NN coupling strength. The scaled covariance $C_{em}$ is defined as $C_{em}=-\frac{L^2}{\langle M^2 \rangle}\left(\langle \varepsilon M^2 \rangle-\langle\varepsilon\rangle \langle M^2\rangle\right)$.
Their leading FSS behaviors are respectively $\chi\sim L^{2-\eta}$ and $C_{em}\sim L^{y_t}$. Including leading correction-to-scaling terms, we extract estimates of $\eta$ and $y_t$ via the following fitting forms:
\begin{align}
    \chi=L^{2-\eta}(a+b L^{-\omega}), \label{chiFit}
\end{align}
and
\begin{align}
    C_{em}=L^{y_t}(a+b L^{-\omega}), \label{KFit}
\end{align}
where $\omega$ is the leading correction exponent. Similar to the procedure used to determine the critical point, we vary the fixed value of $\omega$ to obtain multiple plausible fits. Tables~\ref {yt_exponent} and~\ref {eta_exponent} provide the detailed fitting results for the scaled covariance $C_{em}$ and the susceptibility $\chi$ for different parameter sets, respectively. For $\sigma=1.75,1.875$, and $2$, the final estimates are $\eta\mathrel{=}0.295(5), 0.192(6), 0.132(4)$ and $y_t\mathrel{=}0.34(3), 0.21(2), 0.13(2)$, respectively.

\begin{table}[!h]
    \centering
    \caption{Fitting details for determining the exponent $y_t$ using Eq.~\eqref{KFit}}
    \label{yt_exponent}
    \begin{tabular}{lllllll}
    \hline\hline
    $~\sigma$ & $L_\mathrm{min}$ & $~~~y_t$ & $~~~a$ & $~~b$ & $\omega$ & $\chi^2/\mathrm{DF}$\\ \hline
        1.75 &64 & 0.373(4) & 1.70(5) & -4.9(4) & 0.6 & 3.3/8 \\ 
        &96 & 0.369(7) & 1.76(9) & -5.4(7) &  & 3.0/7 \\ 
        &48 & 0.307(4) & 3.7(1) & -4.7(2) & 0.2 & 2.0/9 \\
        &64 & 0.306(5) & 3.8(1) & -4.8(3) & & 1.9/8 \\ \hline
        1.875&48 & 0.224(7) & 1.39(6) & -9(1) & 1 & 7.8/9 \\
        &64 & 0.22(1) & 1.41(8) & -9(2) & & 7.7/8 \\
        &48 & 0.190(10) & 1.8(1) & -3.2(5) & 0.5 & 7.8/9 \\ 
        &64 & 0.194(15) & 1.8(1) & -3.0(8) & & 7.7/8 \\ \hline
        2&48 & 0.11(2) & 0.9(1) & -4(1) & 0.8 & 9.0/9 \\ 
        &64 & 0.12(2) & 0.9(1) & -3(2) & &8.6/8 \\
        &96 & 0.15(1) & 0.68(4) & - & - & 7.6/8 \\ 
        &128 & 0.15(1) & 0.69(6) & - & - & 7.5/7 \\ \hline
    \end{tabular}
\end{table}

\begin{table}[!h]
    \centering
    \caption{Fitting details for determining the exponent $\eta$ using Eq.~\eqref{chiFit}}
    \label{eta_exponent}
    \begin{tabular}{lllllll}
    \hline\hline
    $~\sigma$ & $L_\mathrm{min}$ & $~~~\eta$ & $~~~a$ & $~~b$ & $\omega$ & $\chi^2/\mathrm{DF}$\\ \hline
        1.75&128 & 0.2999(7) & 0.393(2) & 0.78(2) & 0.6 & 7.1/6 \\
        &192 & 0.2993(8) & 0.391(2) & 0.81(2) & & 5.2/5 \\
        &64 & 0.2909(7) & 0.357(2) & 0.530(7) & 0.45 & 8.9/8 \\
        &128 & 0.291(1) & 0.359(3) & 0.52(1) & & 7.5/6 \\ \hline 
        1.875&48 & 0.1980(2) & 0.3832(9) & 0.457(3) & 0.5 & 9.5/9 \\ 
        &64 & 0.1982(3) & 0.384(1) & 0.454(4) & & 8.2/8 \\ 
        &128 & 0.1873(3) & 0.338(1) & 0.347(3) & 0.35 & 1.9/6 \\ 
        &192 & 0.1877(4) & 0.340(1) & 0.344(4) & & 1.5/5 \\ \hline
        2&48 & 0.1354(2) & 0.5012(9) & 0.386(5) & 0.6 & 8.6/9 \\ 
        &64 & 0.1353(2) & 0.501(1) & 0.392(7) & & 6.9/8 \\ 
        &64 & 0.1289(3) & 0.468(1) & 0.266(4) & 0.4 & 5.4/8 \\ 
        &128 & 0.1292(4) & 0.469(2) & 0.260(6) & & 4.0/6 \\ 
        \hline
    \end{tabular}
\end{table}


\paragraph{Squared magnetization at $\sigma=2$.}
We now present the FSS analysis of the low-temperature properties of the log-QLRO phase. We examine the squared magnetization $\langle M^2\rangle$, the squared correlation-length ratio $(\xi/L)^2$, and the squared Fourier magnetization $\langle M_k^2\rangle$ at $\boldsymbol{k}=(2\pi/L,0)$. The FSS analysis confirms the log-QLRO predictions and shows consistency across these observables.

The spin-wave analysis in the main text predicts
\[
    \langle M^2\rangle\sim[\ln(L/L_0)]^{-\eta_\ell},
\]
where $\eta_\ell$ is the logarithmic decay exponent at a given temperature and $L_0$ is a nonuniversal length scale.
We fit the data to
\begin{equation}
    \langle M^2\rangle=(a+bL^{-\omega})[\ln(L/L_0)]^{-\eta_\ell}+cL^{-2}.
    \label{eq:sm-m2}
\end{equation}
Here, $bL^{-\omega}$ accounts for multiplicative finite-size corrections to the leading logarithmic scaling and $cL^{-2}$ is an additive contribution from analytic backgrounds.
We compare the leading form ($b=c=0$), multiplicative corrections with fixed $\omega=1,2$ ($c=0$), and the additive correction ($b=0$).
We fit each temperature separately, varying $L_{\min}$ from $32$ to $512$.
We assess fit quality using $\chi^2/\mathrm{DF}$, and examine parameter stability as $L_{\min}$ increases.
Table~\ref{tab:fig3-m2-fits} lists representative fits at two adjacent $L_{\min}$ values for each fitting form.
The reported fits use $L_{\max}=4096$ for $\beta=2,4$ and $L_{\max}=2048$ for $\beta=8$.

\begin{table}[!htbp]
\centering
\caption{Fits of $\langle M^2\rangle$ at $\sigma=2$ using Eq.~\eqref{eq:sm-m2}. $L_{\max}=4096$ for $\beta=2,4$ and $2048$ for $\beta=8$.}
\label{tab:fig3-m2-fits}
\begin{tabular}{lllllllll}
\hline\hline
$\beta$ & $L_{\min}$ & $\eta_\ell$ & $a$ & $L_0$ & $b$ & $c$ & $\omega$ & $\chi^2/\mathrm{DF}$ \\
\hline
2 & 128 & $0.356(4)$ & $1.12(1)$ & $0.12(1)$ & -- & -- & -- & $5.0/7$\\
 & 192 & $0.355(5)$ & $1.12(2)$ & $0.12(1)$ & -- & -- & -- & $4.9/6$\\
     & 32 & $0.378(6)$ & $1.21(2)$ & $0.07(1)$ & $0.21(4)$ & -- & $1$ & $10.0/10$\\
 & 48 & $0.373(8)$ & $1.19(3)$ & $0.08(2)$ & $0.17(5)$ & -- & $1$ & $9.0/9$\\
 & 32 & $0.359(3)$ & $1.13(1)$ & $0.113(8)$ & $2.4(4)$ & -- & $2$ & $7.5/10$\\
 & 48 & $0.361(4)$ & $1.14(1)$ & $0.11(1)$ & $3.1(8)$ & -- & $2$ & $6.8/9$\\
 & 32 & $0.360(3)$ & $1.13(1)$ & $0.111(8)$ & -- & $1.3(2)$ & -- & $7.4/10$\\
 & 48 & $0.361(4)$ & $1.14(1)$ & $0.11(1)$ & -- & $1.6(4)$ & -- & $6.9/9$\\
\hline
4 & 48 & $0.1079(6)$ & $0.951(2)$ & $0.40(1)$ & -- & -- & -- & $12.3/10$\\
 & 64 & $0.1085(7)$ & $0.953(2)$ & $0.39(2)$ & -- & -- & -- & $10.6/9$\\
 & 32 & $0.112(2)$ & $0.963(6)$ & $0.30(5)$ & $0.04(1)$ & -- & $1$ & $11.2/10$\\
 & 48 & $0.111(3)$ & $0.960(8)$ & $0.32(6)$ & $0.03(2)$ & -- & $1$ & $10.8/9$\\
 & 32 & $0.1093(9)$ & $0.955(3)$ & $0.37(2)$ & $0.4(2)$ & -- & $2$ & $10.2/10$\\
 & 48 & $0.110(1)$ & $0.956(4)$ & $0.36(3)$ & $0.5(3)$ & -- & $2$ & $10.2/9$\\
 & 32 & $0.109(1)$ & $0.955(3)$ & $0.37(2)$ & -- & $0.3(1)$ & -- & $10.2/10$\\
 & 48 & $0.110(1)$ & $0.956(4)$ & $0.36(3)$ & -- & $0.4(3)$ & -- & $10.2/9$\\
\hline
8 & 32 & $0.0457(3)$ & $0.9673(7)$ & $0.53(2)$ & -- & -- & -- & $12.9/10$\\
 & 48 & $0.0459(3)$ & $0.968(1)$ & $0.51(2)$ & -- & -- & -- & $11.5/9$\\
 & 32 & $0.048(1)$ & $0.974(4)$ & $0.37(9)$ & $0.02(1)$ & -- & $1$ & $10.4/9$\\
 & 48 & $0.048(2)$ & $0.974(6)$ & $0.4(1)$ & $0.02(2)$ & -- & $1$ & $10.4/8$\\
 & 32 & $0.0464(6)$ & $0.969(2)$ & $0.47(4)$ & $0.13(9)$ & -- & $2$ & $10.8/9$\\
 & 48 & $0.0466(8)$ & $0.970(2)$ & $0.46(6)$ & $0.2(2)$ & -- & $2$ & $10.7/8$\\
 & 32 & $0.0464(6)$ & $0.970(2)$ & $0.47(4)$ & -- & $0.13(9)$ & -- & $10.8/9$\\
 & 48 & $0.0466(9)$ & $0.970(2)$ & $0.46(6)$ & -- & $0.2(2)$ & -- & $10.7/8$\\
\hline\hline
\end{tabular}
\end{table}

At $\beta=2$, the leading form gives $\eta_\ell=0.356(4)$ and $0.355(5)$ for $L_{\min}=128$ and $192$, respectively, with consistent estimates of $a$ and $L_0$.
At $\beta=4$ and $8$, it describes the data from $L_{\min}=48$ and $32$, respectively, with exponents near $0.108$ and $0.046$.
The decrease in $\eta_\ell$ with increasing $\beta$ corresponds to slower magnetization decay at lower temperature.

Including finite-size corrections extends the fitting range to smaller sizes.
At all three temperatures, the fits with $\omega=2$ and with additive $L^{-2}$ corrections yield similar estimates of $\eta_\ell$, $a$, and $L_0$.
At $\beta=2$, these fits give $\eta_\ell\simeq0.359$--$0.361$ for $L_{\min}=32,48$, close to the large-size leading-form estimates.
The $\omega=1$ fits yield noticeably larger exponents and smaller $L_0$, indicating strong compensation between the leading and correction terms and suggesting additional systematic uncertainty associated with the assumed correction forms.
At $\beta=4$ and $8$, the corrected fits remain close to the leading-form estimates, with much smaller shifts in $\eta_\ell$ even for $\omega=1$.
We therefore base the central estimates primarily on the large-size leading-form fits and the $\omega=2$ and additive-correction fits, while also accounting for the variation observed in the $\omega=1$ fits when estimating uncertainties.
This leads us to adopt $\eta_\ell=0.36(3)$, $0.11(1)$, and $0.05(1)$ for $\beta=2,4,8$, respectively, with a broader uncertainty at $\beta=2$.

We also fit the squared magnetization to
\begin{equation}
    \langle M^2\rangle=g_0+\frac{b_1}{\ln(L/L_0)}+cL^{-2}.
    \label{M2_fit_eq}
\end{equation}
This form approaches the fitted intercept $g_0$ as $L\to\infty$.
Table~\ref{M2_fit_sigma2} gives positive intercepts near $0.25$, $0.61$, and $0.80$ for $\beta=2,4,8$, respectively.

Both this form and the log-QLRO form fit the data over the available size ranges.
However, a positive fitted intercept does not establish true long-range order.
A slowly decaying logarithmic power can also be approximated by a finite-intercept form over a limited range of sizes.
The log-QLRO form follows from the spin-wave analysis presented above, which predicts $\langle M^2\rangle\sim(\ln L)^{-\eta_\ell}$.
The resulting vanishing magnetization in the thermodynamic limit is also consistent with Bruno's theorem, which rules out finite-temperature long-range order in this model at $\sigma=2$~\cite{Bruno2001}.
Thus, the finite-intercept fits do not necessarily favor true long-range order over log-QLRO.
The apparent saturation can also arise from the extremely slow decay of log-QLRO, making it practically indistinguishable from true LRO as discussed in the main text.

We therefore extend the analysis to $(\xi/L)^2$ and $\langle M_k^2\rangle$, which test complementary predictions of the same spin-wave description.
The logarithmic growth of $(\xi/L)^2$ probes the effective stiffness, while the decay of $\langle M_k^2\rangle$ tests the predicted relation between finite-momentum fluctuations and the magnetization exponent $\eta_\ell$.
Together with the magnetization fits, these tests allow us to assess whether the log-QLRO picture consistently describes the scaling of all three observables.

\begin{table}[!ht]
    \centering
    \caption{Finite-intercept fits of $\langle M^2\rangle$ at $\sigma=2$ using Eq.~\eqref{M2_fit_eq}, for $\beta=2,4,8$. The parameter $g_0$ is the fitted intercept of this form.}
    \label{M2_fit_sigma2}
    \begin{tabular}{lllllll}
    \hline\hline
    $\beta$ & $L_\mathrm{min}$ & $~~~g_0$ & $~~b_1$ & $-\ln L_0$ & $c$ & $\chi^2/\mathrm{DF}$ \\ \hline
        2&96 & 0.253(1) & 3.34(4) & 5.94(7) & - & 4.5/7 \\ 
        &128 & 0.252(2) & 3.37(5) & 6.0(1) & - & 4.1/6 \\
        4&128 & 0.614(2) & 2.06(6) & 7.1(1) & - &  7.2/6 \\ 
        &192 & 0.610(2) & 2.19(6) & 7.5(1) & - &  2.4/5 \\ 
        8&48 & 0.803(1) & 1.15(3) & 7.7(2) & 1.0(1) & 3.4/6 \\
        &64 & 0.802(1) & 1.20(3) & 8.0(2) & 1.4(1) & 1.7/5 \\ \hline
    \end{tabular}
\end{table}


\paragraph{Correlation-length ratio at $\sigma=2$.}
The spin-wave analysis above predicts $(\xi/L)^2\sim s\ln L$ with $s=\kappa_{\rm R}/(N-1)$
[Eqs.~\eqref{eq:SM:xiLfinal} and~\eqref{eq:SM:renorm_params}].
We therefore fit the squared correlation-length ratio to
\begin{equation}
    (\xi/L)^2=a+s\ln L.
    \label{eq:sm-xi2}
\end{equation}
Here, $a$ is the intercept and $s$ is the logarithmic slope, related to the spin stiffness by $\kappa_R=(N-1)s=2s$.
We square the measured $\xi/L$ and propagate its statistical error to first order,
$\delta[(\xi/L)^2]=2(\xi/L)\,\delta(\xi/L)$.
Each temperature is fitted by minimizing $\chi^2$.
The fitting ranges use the same upper cutoffs as the magnetization fits: $L_{\max}=4096$ for $\beta=2,4$ and $2048$ for $\beta=8$.

We vary $L_{\min}$ from $32$ to $512$.
Table~\ref{tab:fig3-xi2} lists the results for $L_{\min}=256,384$.
The slopes change little at $\beta=2,4$, while at $\beta=8$ the slope increases from $3.65(5)$ to $3.74(7)$ and $\chi^2/\mathrm{DF}$ decreases from $6.7/5$ to $2.6/4$.
For comparison, retaining the $\beta=8$, $L=4096$ point at $L_{\min}=256$ gives $s=3.66(5)$ and $\chi^2/\mathrm{DF}=6.8/6$, close to the results with that point excluded. Accounting for statistical errors and systematic uncertainties from the fitting range and finite-size corrections, we adopt the renormalized spin stiffness values $\kappa_{\rm R}=0.90(2)$, $3.00(9)$, and $7.3(3)$ for $\beta=2,4,8$, respectively, consistent with the main text.

\begin{table}[!h]
\centering
\caption{Fits of $(\xi/L)^2=a+s\ln L$ at $\sigma=2$. The fitted slope is $s=\kappa_R/(N-1)$. We use $L_{\max}=4096$ for $\beta=2,4$ and $2048$ for $\beta=8$. Parentheses give statistical errors to one significant digit.}
\label{tab:fig3-xi2}
\begin{tabular}{lllll}
\hline\hline
$\beta$ & $L_{\min}$ & $a$ & $s$ & $\chi^2/\mathrm{DF}$ \\
\hline
2 & 256 & $0.50(3)$ & $0.454(5)$ & $8.1/6$\\
 & 384 & $0.49(4)$ & $0.455(6)$ & $8.1/5$\\
\hline
4 & 256 & $0.0(1)$ & $1.51(2)$ & $8.7/6$\\
 & 384 & $0.1(2)$ & $1.50(2)$ & $8.3/5$\\
\hline
8 & 256 & $-1.5(3)$ & $3.65(5)$ & $6.7/5$\\
 & 384 & $-2.1(5)$ & $3.74(7)$ & $2.6/4$\\
\hline
\hline
\end{tabular}
\end{table}

The marginal spin-wave theory predicts a linear dependence of $s$ (and $\kappa_R$) on $\beta$.
Using the slopes at $L_{\min}=256$, we fit their temperature dependence to
\begin{equation}
    s=a+b\beta.
    \label{eq:sm-xi2-slope}
\end{equation}
The fit gives $a=-0.61(1)$, $b=0.530(6)$, and $\chi^2/\mathrm{DF}=0.23/1$.
Thus, the fitted slopes are consistent with this linear form.

\begin{figure}[!htbp]
    \centering
    \includegraphics[width=0.5\linewidth]{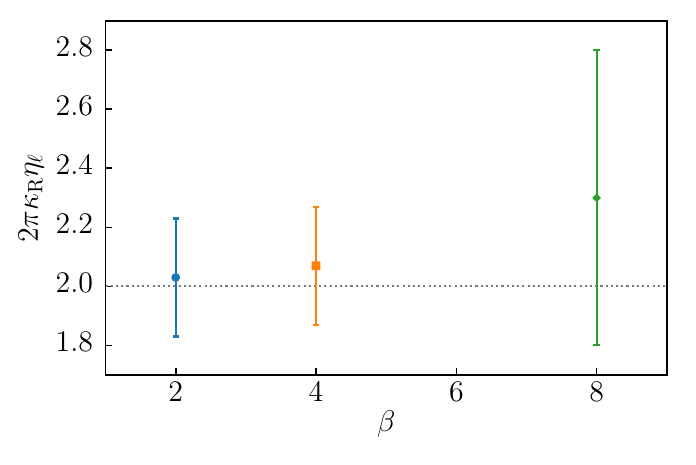}
    \caption{Test of the spin-wave relation
    $\mathcal{P}=2\pi\kappa_{\rm R}\eta_\ell=N-1=2$ at $\sigma=2$.
    The points use the final estimates adopted in the main text:
    $\eta_\ell=0.36(3),0.11(1),0.05(1)$ and
    $\kappa_{\rm R}=0.90(2),3.00(9),7.3(3)$
    for $\beta=2,4,8$, respectively.
    The resulting products are
    $\mathcal{P}=2.03(20),2.07(20),2.30(50)$.
    Error bars show the quoted uncertainties.
    The horizontal dotted line marks $\mathcal{P}=2$.
    All three products are consistent with this prediction.}
    \label{fig:eta_T_identity}
\end{figure}

We then test the identity $\mathcal{P}=2\pi\kappa_{\rm R}\eta_\ell=N-1$ derived in Eq.~\eqref{eq:SM:Pnum}.
Using the final estimates adopted in the main text, $\kappa_{\rm R}=0.90(2),3.00(9),7.3(3)$ and $\eta_\ell=0.36(3),0.11(1),0.05(1)$ for $\beta=2,4,8$, respectively, gives $\mathcal{P}=2.03(20),2.07(20),2.30(50)$.
Figure~\ref{fig:eta_T_identity} shows that all three products are consistent with $N-1=2$ within the quoted uncertainties. 

The agreement with $\mathcal{P}=N-1=2$ provides a nontrivial quantitative test of the spin-wave description of the log-QLRO phase. The slope extracted from $(\xi/L)^2$ and the decay exponent extracted from $\langle M^2\rangle$ satisfy the theory's amplitude relation without an additional adjustable parameter.
This consistency across the three temperatures provides quantitative support for the spin-wave theory as a self-consistent description of the observed low-temperature scaling.

\paragraph{Fourier magnetization at $\sigma=2$.}
The spin-wave theory also predicts the scaling of $\langle M_k^2\rangle$.
As derived in Eq.~\eqref{eq:SM:chikfinal}, the squared Fourier magnetization at the smallest nonzero momentum $k_{\min}=2\pi/L$ has the leading scaling
\begin{equation}
    \langle M_k^2\rangle\sim(\ln L)^{-(1+\eta_\ell)}.
    \label{eq:sm-mk-leading}
\end{equation}
We thus examine this prediction using
\begin{equation}
    \langle M_k^2\rangle=(A+bL^{-\omega})[\ln(L/L_0)]^{-p}.
    \label{eq:sm-mk2}
\end{equation}
The amplitude $A$ and length scale $L_0$ are fitted independently for this observable.
Each temperature is fitted by minimizing $\chi^2$ using the measurement errors, with $L_{\max}=4096$.
We vary $L_{\min}$ from $32$ to $512$.

We first leave $p$ free and fit the leading form with $b=0$.
For $L_{\min}=32,48$, the fitted exponents are $1.04(2),1.03(3)$ at $\beta=2$, $1.04(3),1.03(4)$ at $\beta=4$, and $1.03(3),1.07(5)$ at $\beta=8$, respectively.
We also examine corrections with fixed $\omega=1,2$.
The fits listed in Table~\ref{tab:fig3-mk2-fits} give exponents close to unity.

Since the free-exponent estimates are close to $1$, we next fix $p=1$.
With $b=0$, Eq.~\eqref{eq:sm-mk2} then reduces to $\langle M_k^2\rangle=A/\ln(L/L_0)$.
This form describes the data at all three temperatures, with $\chi^2/\mathrm{DF}$ between about $1.1$ and $1.4$ for the listed fits. At $\beta=2,4$, including the $\omega=2$ correction reduces $\chi^2$ and gives modest changes in $A$ and $L_0$. At $\beta=8$, the correction amplitude is consistent with zero and $\chi^2$ remains unchanged. Because $\eta_\ell$ is small in the low-temperature phase, its contribution to the logarithmic exponent $p=1+\eta_\ell$ is easily absorbed into the scale parameter $L_0$ over the available size range.

\begin{table}[!ht]
\centering
\caption{Fits of $\langle M_k^2\rangle$ at $\sigma=2$ using Eq.~\eqref{eq:sm-mk2}. Parentheses give statistical errors to one significant digit.}
\label{tab:fig3-mk2-fits}
\begin{tabular}{llllllll}
\hline\hline
$\beta$ & $L_{\min}$ & $p$ & $A$ & $L_0$ & $b$ & $\omega$ & $\chi^2/\mathrm{DF}$ \\
\hline
2 & 32 & $1.04(2)$ & $0.025(1)$ & $1.7(1)$ & -- & -- & $12.3/11$\\
 & 48 & $1.03(3)$ & $0.023(2)$ & $1.8(2)$ & -- & -- & $11.4/10$\\
 & 32 & $1.0(1)$ & $0.022(8)$ & $2(2)$ & $-0.01(4)$ & $1$ & $12.2/10$\\
 & 32 & $1.02(5)$ & $0.023(3)$ & $1.9(4)$ & $-0.2(3)$ & $2$ & $11.9/10$\\
 & 32 & $1$ & $0.02202(6)$ & $1.99(2)$ & -- & -- & $17.2/12$\\
 & 48 & $1$ & $0.02194(7)$ & $2.03(3)$ & -- & -- & $12.3/11$\\
 & 32 & $1$ & $0.0217(2)$ & $2.18(9)$ & $-0.015(6)$ & $1$ & $12.2/11$\\
 & 32 & $1$ & $0.02185(9)$ & $2.08(4)$ & $-0.2(1)$ & $2$ & $12.1/11$\\
\hline
4 & 32 & $1.04(3)$ & $0.0129(9)$ & $1.5(2)$ & -- & -- & $13.7/11$\\
 & 48 & $1.03(4)$ & $0.013(1)$ & $1.6(3)$ & -- & -- & $13.5/10$\\
 & 32 & $0.97(5)$ & $0.011(2)$ & $2.2(6)$ & $-0.2(1)$ & $2$ & $11.9/10$\\
 & 32 & $1$ & $0.01168(4)$ & $1.76(3)$ & -- & -- & $16.1/12$\\
 & 48 & $1$ & $0.01164(5)$ & $1.79(3)$ & -- & -- & $14.2/11$\\
 & 32 & $1$ & $0.01158(6)$ & $1.85(5)$ & $-0.14(7)$ & $2$ & $12.2/11$\\
\hline
8 & 32 & $1.03(3)$ & $0.0064(5)$ & $1.5(2)$ & -- & -- & $13.9/11$\\
 & 48 & $1.07(5)$ & $0.0072(9)$ & $1.2(3)$ & -- & -- & $12.2/10$\\
 & 32 & $1$ & $0.00602(2)$ & $1.64(3)$ & -- & -- & $14.5/12$\\
 & 48 & $1$ & $0.00602(3)$ & $1.64(4)$ & -- & -- & $14.4/11$\\
 & 32 & $1$ & $0.00602(4)$ & $1.64(6)$ & $0.00(4)$ & $2$ & $14.5/11$\\
\hline\hline
\end{tabular}
\end{table}

We next test the prediction $p=1+\eta_\ell$ using the central magnetization estimates adopted in the main text.
We fix $p=1.36$, $1.11$, and $1.05$ for $\beta=2,4,8$, respectively, while keeping the amplitude $A$ and scale $L_0$ free.
At $\beta=2,4$, we compare the leading form with fits that include a finite-size correction. At $\beta=8$, we use only the leading form. Table~\ref{tab:fig3-mk2-eta} lists the results.

At $\beta=2,4$, the leading form describes the large-size data at $L_{\min}=256,384$.
The estimates of $A$ and $L_0$ remain consistent for these two values of $L_{\min}$.
Including the $\omega=1$ correction extends the fitting range to $L_{\min}=48$ at $\beta=2$ and $64$ at $\beta=4$.
At $\beta=2$, the correction amplitudes are $b=0.24(3)$ and $0.24(4)$ for $L_{\min}=48\mathpunct{,}64$.
At $\beta=4$, they are $b=0.03(1)$ and $0.05(2)$ for $L_{\min}=64\mathpunct{,}96$.
At $\beta=8$, the leading form already gives good fits at $L_{\min}=32,48$, so no correction term is needed for these fits.

With $p$ fixed by the magnetization analysis, the fits describe the $\langle M^2_{k}\rangle$ data at all three temperatures.
This provides a further quantitative consistency check of the spin-wave prediction.

\begin{table}[!ht]
\centering
\caption{Fits of $\langle M_k^2\rangle$ at $\sigma=2$ using
Eq.~\eqref{eq:sm-mk2}, with $p=1+\eta_\ell$ fixed by the estimates adopted in the main text. Parentheses give statistical errors to one significant digit.}
\label{tab:fig3-mk2-eta}
\begin{tabular}{llllllll}
\hline\hline
$\beta$ & $L_{\min}$ & $p$ & $A$ & $L_0$ & $b$ & $\omega$ & $\chi^2/\mathrm{DF}$ \\
\hline
2 & 256 & $1.36$ & $0.0631(7)$ & $0.25(2)$ & -- & -- & $8.3/6$\\
  & 384 & $1.36$ & $0.0630(9)$ & $0.25(2)$ & -- & -- & $8.2/5$\\
  & 48  & $1.36$ & $0.0648(8)$ & $0.20(2)$ & $0.24(3)$ & $1$ & $12.2/10$\\
  & 64  & $1.36$ & $0.0648(9)$ & $0.20(2)$ & $0.24(4)$ & $1$ & $12.2/9$\\
\hline
4 & 256 & $1.11$ & $0.0160(2)$ & $0.88(6)$ & -- & -- & $8.8/6$\\
  & 384 & $1.11$ & $0.0161(2)$ & $0.85(8)$ & -- & -- & $8.4/5$\\
  & 64  & $1.11$ & $0.0161(2)$ & $0.83(8)$ & $0.03(1)$ & $1$ & $11.3/9$\\
  & 96  & $1.11$ & $0.0163(3)$ & $0.8(1)$  & $0.05(2)$ & $1$ & $10.7/8$\\
\hline
8 & 32 & $1.05$ & $0.00684(3)$ & $1.31(2)$ & -- & -- & $14.4/12$\\
  & 48 & $1.05$ & $0.00687(3)$ & $1.28(3)$ & -- & -- & $12.3/11$\\
\hline\hline
\end{tabular}
\end{table}


\paragraph{Goldstone modes and long-range order for $\sigma<2$.}
In this part, we present a finite-size scaling analysis of the low-temperature Goldstone modes for $\sigma<2$.
In this regime, the low-temperature phase has true long-range order, as discussed in the main text.
The squared magnetization therefore approaches a nonzero thermodynamic-limit value, with its leading finite-size correction set by the transverse Goldstone modes.
For $0<\sigma<2$, the spin-wave kernel $\gamma(k)\simeq A_\sigma k^\sigma$ in Eq.~\eqref{eq:SM:kernels} predicts a correction governed by the decay exponent $2-\sigma$ in two dimensions:
\[
    \langle M^2\rangle
    =g_0+b_1L^{\sigma-2}+\cdots,
    \qquad g_0>0.
\]

For $\sigma=1.75$ and $1.875$, we fit the squared magnetization to
\begin{equation}
    \langle M^2\rangle
    =g_0+L^{\sigma-2}(b_1+b_2L^{-\omega})+cL^{-2}.
    \label{eq:sm-m2-lro}
\end{equation}
Here, $g_0$ is the thermodynamic-limit value of $\langle M^2\rangle$.
The term proportional to $b_2$ describes a subleading finite-size correction, and $cL^{-2}$ is an analytic contribution.
Table~\ref{M2_fit_l2} lists the fits at $\beta=2$, with $c=0$ and several fixed values of $\omega$.
For each fixed $\omega$, the fitted parameters are stable as $L_{\min}$ increases.
All listed fits give $g_0>0$, with values near $0.55$ for $\sigma=1.75$ and $0.46$ for $\sigma=1.875$.
The nonzero intercepts across the tested correction exponents support the true-LRO interpretation in the main text, although the precise extrapolated values depend on the correction form.

\begin{table}[!ht]
    \centering
    \caption{Fits of $\langle M^2\rangle$ for $\sigma<2$ at $\beta=2$ using Eq.~\eqref{eq:sm-m2-lro}, with $c=0$ and fixed $\omega$.}
    \label{M2_fit_l2}
    \begin{tabular}{lllllll}
    \hline\hline
        $~\sigma$ & $L_\mathrm{min}$ & $~~~~g_0$ & $~~~b_1$ & $~~~b_2$ & $\omega$ & $\chi^2/\mathrm{DF}$ \\ \hline
        1.75 & 96 & 0.5449(1) & 0.164(1) & 0.121(3) & 0.3 & 6.5/7 \\ 
        &128 & 0.5450(1) & 0.162(1) & 0.125(3) & & 3.8/6 \\
        &96 & 0.5478(1) & 0.076(2) & 0.172(3)&0.1 & 2.7/7 \\
        &128 & 0.5479(1) & 0.075(3) & 0.173(4)& & 2.5/6 \\ \hline
        1.875 & 64 & 0.4573(3) & 0.190(1) & 0.186(1) & 0.32  & 8.2/8 \\
        &96 & 0.4574(4) & 0.190(1) & 0.187(2) &  & 8.0/7 \\
        &64 & 0.4637(4) & 0.162(1) & 0.189(2) & 0.25 & 8.8/8 \\
        &96 & 0.4633(5) & 0.163(1) & 0.187(2) & & 7.4/7 \\ \hline
    \end{tabular}
\end{table}

At nonzero momentum, the uniform LRO contribution is absent, so the leading signal comes from the Goldstone fluctuations.
We therefore fit the squared Fourier magnetization at the minimum nonzero momentum to
\begin{equation}
    \langle M_k^2\rangle=L^{\sigma-2}(a_0+a_1L^{-\omega}).
    \label{Mk2_fit_eq}
\end{equation}
Here, $a_0$ is the leading amplitude, and $a_1L^{-\omega}$ describes its finite-size correction.
Table~\ref{Mk2_fit_sigmal2} lists the fits at $\beta=2$.
For each $\sigma$, the two fitting ranges give consistent amplitudes and correction exponents.

Taken together, the positive extrapolated value of $\langle M^2\rangle$ and the $L^{\sigma-2}$ scaling of $\langle M_k^2\rangle$ are consistent with spontaneous symmetry breaking and the associated Goldstone modes for $\sigma<2$~\cite{Yao2025LRXY}.

\begin{table}[!ht]
    \centering
    \caption{Fits of $\langle M_k^2\rangle$ for $\sigma<2$ at $\beta=2$ using Eq.~\eqref{Mk2_fit_eq}.}
    \label{Mk2_fit_sigmal2}
    \begin{tabular}{llllll}
    \hline\hline
    $\sigma$ & $L_\mathrm{min}$ & $~~a_0$ & $~~a_1$ & $\omega$ & $\chi^2/\mathrm{DF}$ \\ \hline
        
        1.75&48 & 0.00875(5) & 0.034(1) & 0.74(1) & 5.9/9 \\
        &64 & 0.00867(6) & 0.031(1) & 0.72(1) & 3.9/8 \\ 
        1.875&96 & 0.0047(1) & 0.027(2) & 0.56(2) & 6.2/5 \\
        &128 & 0.0048(1) & 0.031(3) & 0.59(3) & 4.2/4 \\ 
        \hline
    \end{tabular}
\end{table}

\bibliography{ref_fix}